\newif\ifdraft
\draftfalse
\newif\ifcameraready
\camerareadytrue

\documentclass[camera,letterpaper,nomarginnotes,nonarrowgutter]{jpaper}

\usepackage[bookmarks=true,breaklinks=true,colorlinks,linkcolor=black,citecolor=NavyBlue,urlcolor=blue]{hyperref}

\usepackage{algorithmic}
\usepackage{graphicx}
\usepackage{textcomp}
\usepackage{xcolor}
\usepackage[flushleft]{threeparttable}

\usepackage{titlecaps}

\usepackage{balance}
\usepackage{textcomp}
\usepackage{pifont}
\usepackage{booktabs}
\usepackage{glossaries-extra} % For acronyms / abbreviations
\setkeys{glslink}{hyper=false} % to disable hyperlink for glossaries
\usepackage{setspace}
\usepackage{listings}
\usepackage[linesnumbered]{algorithm2e}
\usepackage{siunitx} % To use SI unit system 
\usepackage{tikz}
\usepackage[dvipsnames]{xcolor}
\usepackage{multirow}
\usepackage{enumitem}
\usepackage[utf8]{inputenc}

\usepackage{titlesec}

\usepackage{interval} %sec5, interval
\usepackage{duckuments} %duck figures
\usepackage{algorithmic} %algoritmic

\usepackage[us,12hr]{datetime}
\usepackage[en-GB, useregional=numeric]{datetime2}
\usepackage{fourier-orns}
\usepackage{fancyhdr}
\usepackage{pgfornament}
\usetikzlibrary{calc}
\usepackage{url}
\usepackage{yfonts}
\usepackage[sort,compress]{cite}

\usepackage{amsmath,amssymb,amsfonts}

\usepackage{mathptmx} % This is Times font
\usepackage{subcaption}

\def\UrlBreaks{\do\/\do-\/\do.\/\do:}

\expandafter\def\expandafter\UrlBreaks\expandafter{\UrlBreaks
  \do\a\do\b\do\c\do\d\do\e\do\f\do\g\do\h\do\i\do\j
  \do\k\do\l\do\m\do\n\do\o\do\p\do\q\do\r\do\s\do\t
  \do\u\do\v\do\w\do\x\do\y\do\z\do\A\do\B\do\C\do\D
  \do\E\do\F\do\G\do\H\do\I\do\J\do\K\do\L\do\M\do\N
  \do\O\do\P\do\Q\do\R\do\S\do\T\do\U\do\V\do\W\do\X
  \do\Y\do\Z}

\newcommand{\revdel}[1]{}

\newcommand*\DRAMTIMING[1]{t\textsubscript{#1}}

\newcommand*\nCHIPS{216}
\newcommand*\nMODULES{28}

\definecolor{ao}{rgb}{0.007, 0.520, 0.867}
\definecolor{brown(web)}{rgb}{0.65, 0.16, 0.16}
\definecolor{bisque}{rgb}{1.0, 0.89, 0.77}
\definecolor{nbs}{rgb}{0.88, 0.07, 0.37}
\definecolor{iy}{rgb}{0.0, 0.56, 0.041}
\newcommand{\one}{1)}
\newcommand{\two}{2)}
\newcommand{\three}{3)}

\newcommand{\param}[1]{\textcolor{red}{#1}}

\ifdraft
    \usepackage[colorinlistoftodos,prependcaption,textsize=scriptsize]{todonotes} % For margin notes
    \paperwidth=\dimexpr \paperwidth + 4cm\relax
    \oddsidemargin=\dimexpr\oddsidemargin + 2cm\relax
    \evensidemargin=\dimexpr\evensidemargin + 2cm\relax
    \marginparwidth=\dimexpr \marginparwidth + 2cm\relax
    \newcommand{\agy}[1]{\textcolor{gfored}{#1}}
    \newcommand{\agycomment}[1]{\todo[size=\scriptsize, linecolor=orange, bordercolor=orange, backgroundcolor=white]{\textcolor{gfored}{\textbf{@gy:} #1}}}

    \newcommand{\atb}[1]{\textcolor{ao}{#1}}

    \newcommand{\atbcomment}[1]{\todo[size=\scriptsize, linecolor=orange, bordercolor=orange, backgroundcolor=white]{\textcolor{ao}{\textbf{@atb:} #1}}}

    \newcommand{\nb}[1]{\textcolor{nbs}{#1}}
    \newcommand{\nbcomment}[1]{\todo[size=\scriptsize, linecolor=orange, bordercolor=orange, backgroundcolor=white]{\textcolor{nbs}{\textbf{@nb:} #1}}}

    \newcommand{\hluo}[1]{\textcolor{moegi}{#1}}
    \newcommand{\hluocomment}[1]{\todo[size=\scriptsize, linecolor=orange, bordercolor=orange, backgroundcolor=white]{\textcolor{moegi}{\textbf{@hluo:} #1}}}

    \newcommand{\revcomment}[1]{\todo[size=\scriptsize, linecolor=orange, bordercolor=orange, backgroundcolor=white]{\textcolor{red}{\textbf{Full Comment:} #1}}}

    \newcommand{\iey}[1]{\textcolor{iy}{#1}}
    \newcommand{\ieycomment}[1]{\todo[size=\scriptsize, linecolor=orange, bordercolor=orange, backgroundcolor=white]{\textcolor{iy}{\textbf{@iey:} #1}}}

    \newcommand{\om}[1]{\textcolor{blue}{#1}}
    \newcommand{\omcomment}[1]{\todo[size=\scriptsize, linecolor=orange, bordercolor=orange, backgroundcolor=white]{\textcolor{blue}{\textbf{@om:} #1}}}

\else
    \renewcommand{\param}[1]{\textcolor{black}{#1}}

    \newcommand{\agy}[1]{{#1}}
    \newcommand{\agycomment}[1]{}
    \newcommand{\agyinline}[1]{}

    \newcommand{\atb}[1]{{#1}}
    \newcommand{\atbcomment}[1]{}

    \newcommand{\nb}[1]{{#1}}
    \newcommand{\nbcomment}[1]{}

    \newcommand{\hluo}[1]{{#1}}
    \newcommand{\hluocomment}[1]{}

    \newcommand{\iey}[1]{#1}
    \newcommand{\ieycomment}[1]{}

    \newcommand{\revcomment}[1]{}

    \newcommand{\om}[1]{#1}
    \newcommand{\omcomment}[1]{}

\fi

\definecolor{frenchblue}{rgb}{0.19, 0.55, 0.91}

\usepackage[most]{tcolorbox} 
\tcbset{before skip=4.1pt, after skip=4.1pt}

\newtcolorbox[auto counter]{obsx}[3][]{%
    colframe = #2!45,
    colback  = #2!10,
    coltitle = #2!20!black, 
    colbacktitle=#2!20,
    coltitle=black,
    fonttitle=\bfseries, 
    title=#3~\thetcbcounter.\ ,
    enhanced,
    attach boxed title to top left={yshift=-2.8mm, xshift=0.15cm},
    bottom=-2.2pt,
    #1% 
}

\usepackage[most]{tcolorbox} 
\newtcolorbox[auto counter]{tkx}[2][]{%
    enhanced, breakable, center title,
    colframe = #2!45,
    colback  = #2!10,
    colbacktitle=#2!20,
    left=-0.5pt,
    right=-0.5pt,
    bottom=-2pt,
    top=-0.25pt,
    #1% 
}
\newcounter{obs}
\newcommand\observation[1]{
\refstepcounter{obs}
\begin{tkx}{Plum}
\noindent\textbf{Observation~\theobs.} #1
\end{tkx}
}

\newcounter{tkw}
\newcommand\takeaway[1]{
\stepcounter{tkw}
\begin{tkx}{Sepia}
\noindent\textbf{Takeaway~\thetkw.} #1
\end{tkx}
}

\newcounter{hypo}
\newcommand\hypothesis[1]{
\refstepcounter{hypo}
\begin{tkx}{NavyBlue}
\noindent\textbf{Key Hypothesis.} #1
\end{tkx}
}

\newcommand{\figref}[1]{Fig.~\ref{#1}}

\newcommand{\secref}[1]{§\ref{#1}}

\ifcameraready

    \newcommand{\atbcr}[2]{\ifnum#1>-1\textcolor{black}{#2}\else{#2}\fi}
    \newcommand{\ieycr}[2]{\ifnum#1>-1\textcolor{black}{#2}\else{#2}\fi}
    \newcommand{\omcr}[2]{\ifnum#1>-1\textcolor{black}{#2}\else{#2}\fi}
    \newcommand{\omcrcomment}[1]{}
    \newcommand{\crdiscussion}[2]{}{}

    \newcommand{\ominline}[1]{}
    \newcommand{\ieycrcomment}[1]{}
    \newcommand{\atbcrcomment}[1]{}
    \newcommand{\agycrcomment}[1]{}
    \newcommand{\ieyinline}[1]{}
\else
\usepackage[colorinlistoftodos,prependcaption,textsize=scriptsize]{todonotes} % For margin notes
    \paperwidth=\dimexpr \paperwidth + 4cm\relax
    \oddsidemargin=\dimexpr\oddsidemargin + 2cm\relax
    \evensidemargin=\dimexpr\evensidemargin + 2cm\relax
    \marginparwidth=\dimexpr \marginparwidth + 2cm\relax

    \newcommand{\atbcr}[2]{\ifnum#1=\value{version}\textcolor{ao}{#2}\else{#2}\fi}

    \newcommand{\ieycr}[2]{\ifnum#1=\value{version}\textcolor{blue}{#2}\else{#2}\fi}

    \newcommand{\ieycrcomment}[1]{\todo[linecolor=orange, bordercolor=orange, backgroundcolor=white]{\textcolor{iy}{\textbf{@Ismail:} #1}}}
    \newcommand{\ieyinline}[1]{\\\textcolor{iy}{\textbf{@Ismail:} #1}}

    \newcommand{\atbcrcomment}[1]{\todo[linecolor=brown, bordercolor=brown, backgroundcolor=white]{\textcolor{ao}{\textbf{@Atb:} #1}}}

    \newcommand{\agycrcomment}[1]{\todo[size=\scriptsize, linecolor=orange, bordercolor=orange, backgroundcolor=white]{\textcolor{gfored}{\textbf{@gy:} #1}}}
    
    \newcommand{\crdiscussion}[2]{\omcrcomment{#1\\\textcolor{blue}{\textbf{@Ismail:}#2}}}
    \newcommand{\omcr}[2]{\ifnum#1=\value{version}\textcolor{red}{#2}\else{#2}\fi}

    \newcommand{\omcrcomment}[1]{\todo[linecolor=orange, bordercolor=orange, backgroundcolor=white]{\textcolor{red}{\textbf{@Onur:} #1}}}
    \newcommand{\ominline}[1]{\\\textcolor{red}{\textbf{@Onur:} #1}}

\fi

\makeatletter
\g@addto@macro{\normalsize}{%
  \setlength{\abovedisplayskip}{4pt plus 0.5pt minus 1pt}
  \setlength{\belowdisplayskip}{3pt plus 0.5pt minus 1pt}
  \setlength{\abovedisplayshortskip}{0pt}
  \setlength{\belowdisplayshortskip}{0pt}
  \setlength{\intextsep}{5pt plus 1pt minus 1pt}
  \setlength{\textfloatsep}{3pt plus 1pt minus 1pt}
  \setlength{\skip\footins}{5pt plus 1pt minus 1pt}
  \setlength{\abovecaptionskip}{2pt plus 0pt minus 0pt}}
\makeatother

\setabbreviationstyle[acronym]{long-short}

\newcommand{\X}[0]{ColumnDisturb}

\newcommand{\tras}[0]{t_{RAS}}
\newcommand{\trp}[0]{t_{RP}}

\newcommand{\trefi}[0]{t_{REFI}}
\newcommand{\trefw}[0]{t_{REFW}}

\newcommand{\pum}[0]{PuM}
\newcommand{\pim}[0]{PiM}
\newcommand{\pnm}[0]{PnM}

\newcommand{\pud}[0]{{PuD}}

\newcommand{\act}[0]{\texttt{ACT}}
\newcommand{\pre}[0]{\texttt{PRE}}
\newcommand{\refresh}[0]{REF}
\newcommand{\wri}[0]{\texttt{WR}}
\newcommand{\rd}[0]{\texttt{RD}}

\newcommand{\cots}[0]{COTS}
\newcommand{\comra}[0]{CoMRA}
\newcommand{\simra}[0]{SiMRA}

\newacronym{iqr}{$IQR$}{inter-quartile range}
\newacronym{act}{\act{}}{activate}
\newacronym{pre}{\pre{}}{precharge}
\newacronym{ref}{\refresh{}}{refresh}
\newacronym{wr}{\wri{}}{write}
\newacronym{rd}{\rd{}}{read}
\newacronym{pim}{\pim{}}{Processing-in-Memory}
\newacronym{pnm}{\pnm{}}{Processing-near-Memory}

\newacronym{pum}{\pum{}}{Processing-using-Memory}
\newacronym{pud}{\pud{}}{Processing-using-DRAM}

\newacronym{cots}{\cots{}}{commercial off-the-shelf}
\newacronym{comra}{\comra{}}{consecutive multiple-row activation}
\newacronym{simra}{\simra{}}{simultaneous multiple-row activation}

\newacronym{jedec}{JEDEC}{Joint Electron Device Engineering Council}
\newcommand{\hcfirst}[0]{$HC_{first}$}
\newacronym{hcfirst}{\hcfirst{}}{the minimum {hammer count} {required to induce the first bitflip}} 
\newcommand{\ber}[0]{$BER$}
\newacronym{ber}{\ber{}}{bit error rate}
\newacronym{wcdp}{$WCDP$}{worst-case data pattern}
\newacronym{taggon}{$t_{AggOn}$}{the time that an aggressor row stays active}

\newacronym{trefw}{$\trefw$}{}
\newacronym{tras}{$\tras$}{}
\newacronym{trp}{$\trp$}{}
\newacronym{trefi}{$\trefi$}{}

\newcommand{\exploitingRowHammerAllCitations}[0]{\cite{fournaris2017exploiting, poddebniak2018attacking, tatar2018throwhammer, carre2018openssl, barenghi2018software, zhang2018triggering, bhattacharya2018advanced, google-project-zero, kim2014flipping, rowhammergithub, seaborn2015exploiting, van2016drammer, gruss2016rowhammer, razavi2016flip, pessl2016drama, xiao2016one, bosman2016dedup, bhattacharya2016curious, burleson2016invited, qiao2016new, brasser2017can, jang2017sgx, aga2017good, mutlu2017rowhammer, tatar2018defeating, gruss2018another, lipp2018nethammer, van2018guardion, frigo2018grand, cojocar2019eccploit,  ji2019pinpoint, mutlu2019rowhammer, hong2019terminal, kwong2020rambleed, frigo2020trrespass, cojocar2020rowhammer, weissman2020jackhammer, zhang2020pthammer, yao2020deephammer, deridder2021smash, hassan2021utrr, jattke2022blacksmith, tol2022toward, kogler2022half, orosa2022spyhammer, zhang2022implicit, liu2022generating, cohen2022hammerscope, zheng2022trojvit, fahr2022frodo, tobah2022spechammer, rakin2022deepsteal, aydin2022cyber, mus2022jolt, wang2022research, lefforge2023reverse,fahr2022effects, kaur2022work, cai2022feasibility, li2022cyberradar, roohi2022efficient, staudigl2022neurohammer, yang2022socially, islam2022signature}}

\newcommand{\mitigatingRowHammerAllCitations}[0]{\cite{AppleRefInc, rh-hp,rh-lenovo,greenfield2012throttling, kim2014flipping, kim2014architectural, bains14d, bains14c, bains14a, bains14b, aweke2016anvil, bains2015row, bains2016row, bains2016distributed, son2017making, seyedzadeh2018cbt,irazoqui2016mascat, you2019mrloc, lee2019twice, park2020graphene, yaglikci2021security, yaglikci2021blockhammer, frigo2020trrespass, kang2020cattwo, hassan2021utrr, qureshi2022hydra, saileshwar2022randomized, brasser2017can, konoth2018zebram, van2018guardion, vig2018rapid,  kim2022mithril, lee2021cryoguard, marazzi2022protrr, zhang2022softtrr, joardar2022learning, juffinger2023csi, yaglikci2022hira, saxena2022aqua, enomoto2022efficient, manzhosov2022revisiting, ajorpaz2022evax, naseredini2022alarm, joardar2022machine,  zhang2020leveraging,loughlin2021stop, devaux2021method, fakhrzadehgan2022safeguard, saroiu2022price, loughlin2022moesiprime, han2021surround, mutlu2023fundamentally, woo2023scalable, bock2019riprh, kim2015architectural, wang2021discreet, bennett2021panopticon, olgun2024abacus, bostanci2024comet, canpolat2024understanding, canpolat2025chronus, tugrul2025understanding,yaglikci2024svard,taneja2025dream,vittal2025mopac,lin2025cnc,qazi2025drfm,lu2025counterpoint,qureshi2025autorfm,woo2025dapper,woo2025qprac,bostanci2025understanding}}

\newcommand{\retentionMechanisms}[0]{\cite{lin2012secret,liu2012raidr,nair2013arch,ohsawa1998optimizing,venkatesan2006rapid,wang2014proactivedram,baek2014refresh,bhati2015dram,cui2014dtail,khan2014efficacy,liu2013experimental,qureshi2015avatar,wang2015radar,liu2011flikker}}

\begin{document}

\title{{
ColumnDisturb: Understanding Column-based Read Disturbance in Real DRAM Chips and Implications for Future Systems
}}

\newcommand{\affilETH}[0]{\textsuperscript{1}}
\newcommand{\affilCISPA}[0]{\textsuperscript{2}}
\author{
{{\.I}smail Emir Y{\"u}ksel\affilETH}\qquad
{Ataberk Olgun\affilETH}\qquad
{F. Nisa Bostanc{\i}\affilETH}\\
{Haocong Luo\affilETH}\qquad
{A.~Giray~Ya\u{g}l{\i}k\c{c}{\i}\affilETH\textsuperscript{,}\affilCISPA}\qquad
{Onur Mutlu\affilETH}
\vspace{0mm}\\
{\affilETH ETH Z{\"u}rich\qquad{} \affilCISPA CISPA}
}
%%%%%%%%%%%%%%%%%%%%%%%%%%%%%%%%%%%%

\maketitle

    \renewcommand{\headrulewidth}{0pt}
    \fancypagestyle{firstpage}{
        \fancyhead{} 
        \fancyhead[C]{
      } 
    \renewcommand{\footrulewidth}{0pt}
    }
  \thispagestyle{firstpage}

\pagenumbering{arabic}

\newcounter{version}
\setcounter{version}{999}
\glsresetall
\setstretch{0.925}
\begin{abstract}
\omcr{0}{W}e experimentally demonstrate a new widespread read disturbance phenomenon, \X{}, in real \om{commodity} DRAM chips. 
By repeatedly opening or keeping a DRAM row (aggressor row) open, we show that it is possible to disturb DRAM cells through \omcr{0}{a} \emph{DRAM column} \omcr{0}{(i.e., bitline)} and induce bitflips in DRAM cells sharing the same columns 
as the aggressor row\ieycr{0}{ \omcr{2}{(across multiple DRAM subarrays)}. \omcr{1}{With} \X{}}\omcr{1}{, the activation of a single row}
\om{concurrently} \omcr{2}{disturbs} \ieycr{0}{DRAM} cells across as many as \emph{three DRAM subarrays} (e.g., up to 3072 {DRAM} rows in \iey{tested} \agy{DRAM} chip\iey{s}) \omcr{1}{as opposed to RowHammer \& RowPress, which affect only a few neighboring rows of \omcr{2}{the} aggressor row in a single subarray}.  
We rigorously and comprehensively characterize \X{} and its \omcr{0}{characteristics} under various operational conditions (i.e., temperature, data pattern, \ieycr{0}{DRAM timing parameters, average voltage level of the bitline, memory access pattern, and spatial variation}) using \nCHIPS{} DDR4 \iey{and 4 HBM2 chips from three major {DRAM} manufacturers}. Among our \om{27} key \omcr{0}{experimental} observations, we highlight two \om{major} results and their implications.

First, \X{} affects chips from all three major DRAM manufacturers and worsens as DRAM technology scales down to smaller node sizes (e.g., the minimum time to induce the first \X{} bitflip reduces by up to \param{5.06}x \ieycr{1}{and \param{2.96}x on average across all tested DRAM modules)}. We observe that, even in existing DRAM chips, \X{} induces bitflips within a standard DDR4 refresh {window} (e.g., in 63.6 ms) in multiple cells from one module. \om{We predict that}, \omcr{0}{as DRAM technology node size reduces,} \X{} would worsen in future DRAM chips, \om{likely} causing many more bitflips in the standard refresh {window}. 
\atb{Second, \ieycr{0}{beyond} the standard refresh {window}, \X{} induces \ieycr{0}{bitflips in many \omcr{1}{(up to \param{198}x)} more DRAM rows than retention failures.}
There\omcr{0}{fore}, \X{} \atb{has strong implications for existing retention-aware
refresh mechanisms that aim to improve system performance and energy
efficiency by leveraging the heterogeneity in DRAM cell retention time\omcr{1}{s}:
our detailed analyses and cycle-level simulations show that 
\X{} greatly reduces and can completely diminish the performance and 
energy benefits of such mechanisms.}}

\iey{O}ur results have serious implications for the robustness of future \omcr{0}{DRAM-based computing} systems due to continued aggressive DRAM technology scaling. \omcr{0}{W}e \omcr{0}{describe} \atb{and} \omcr{0}{evaluate} \param{two} solutions \ieycr{0}{for mitigating \X{} bitflips \omcr{1}{and call for more research on the topic}.}

\end{abstract}
\glsresetall
\section{Introduction}
\label{sec:intro}
Dynamic random access memory (DRAM)~\cite{dennard1968dram} is the dominant main memory technology \omcr{0}{in} nearly all modern computing systems due to its \omcr{0}{latency and} cost \omcr{0}{characteristics}. Continuing to increase DRAM capacity requires increasing the density of DRAM cells by reducing (i.e., scaling) the technology node size of DRAM, which in turn reduces DRAM cell size, cell-to-cell spacing, and cell-to-bitline spacing~\cite{Yu2022TheStudy}. Unfortunately, as a result of this aggressive technology node scaling, DRAM suffers from worsening read disturbance \omcr{1}{problems}~\cite{kim2020revisiting,luo2023rowpress,mutlu2019rowhammer,mutlu2023fundamentally}. 

Read disturbance in modern DRAM chips~\cite{kim2014flipping,
mutlu2019rowhammer,mutlu2023fundamentally,luo2023rowpress, olgun2024read,
mutlu2017rowhammer} is a widespread phenomen\omcr{0}{on} and is reliably used for breaking memory isolation~\exploitingRowHammerAllCitations{}, a fundamental building block \iey{of} robust \omcr{0}{(i.e., safe, secure, reliable, available)} systems. \emph{RowHammer} and \emph{RowPress} are two prominent examples of {DRAM read disturbance phenomena {where} a victim DRAM row
experiences bitflips when a nearby aggressor DRAM row is} 1)~repeatedly activated (i.e., hammered)~\cite{kim2014flipping,mutlu2019rowhammer,mutlu2023fundamentally} or 2)~kept open for a long period (i.e., pressed)~\cite{orosa2021deeper, luo2023rowpress}. Many prior works~\cite{mutlu2017rowhammer, mutlu2019rowhammer, frigo2020trrespass, cojocar2020rowhammer, kim2020revisiting, kim2014flipping, luo2023rowpress} experimentally demonstrate that read disturbance significantly
worsens as DRAM manufacturing technology scales to smaller node \omcr{0}{sizes in real DRAM chips}. For example, chips manufactured in 2018-2020 can experience RowHammer bitflips at an order of magnitude fewer row activations compared to the chips manufactured in 2012-2013~\cite{kim2020revisiting}.
To ensure the robustness of modern and future DRAM-based systems, it is critical to \omcr{0}{1)} develop a rigorous understanding of \omcr{1}{\emph{known}} read disturbance effects like RowHammer and RowPress \agy{and \omcr{0}{2)} discover \omcr{1}{and analyze \emph{previously unknown}} read disturbance phenomena}.

In this work, for the first time, we experimentally demonstrate \omcr{0}{a new} widespread read disturbance phenomenon, \X{}, in \nCHIPS{} \gls{cots} DDR4 \iey{and 4 HBM2} DRAM chips from all three major \omcr{0}{DRAM} manufacturers. We show that it is possible to \atb{disturb} many DRAM cells that are on the same \emph{DRAM column}.\ieycr{0}{\footnote{In DRAM, each \emph{logical} column is implemented using multiple physical columns, i.e., bitlines \ieycr{0}{(e.g.,~\cite{micron2014ddr4,hynixh8,samsung-real-datasheet})}. In this paper, we use the term \emph{DRAM column} to refer to a single physical column/bitline.}} \atb{\X{} manifests as bitflips in
DRAM cells across as many as three DRAM subarrays (e.g., up to 3072 DRAM rows in \iey{tested} DDR4 chip\iey{s}) \emph{concurrently} when we hammer or press
an aggressor row.}

\begin{figure}[ht]
    \centering
    \includegraphics[width=1\linewidth]{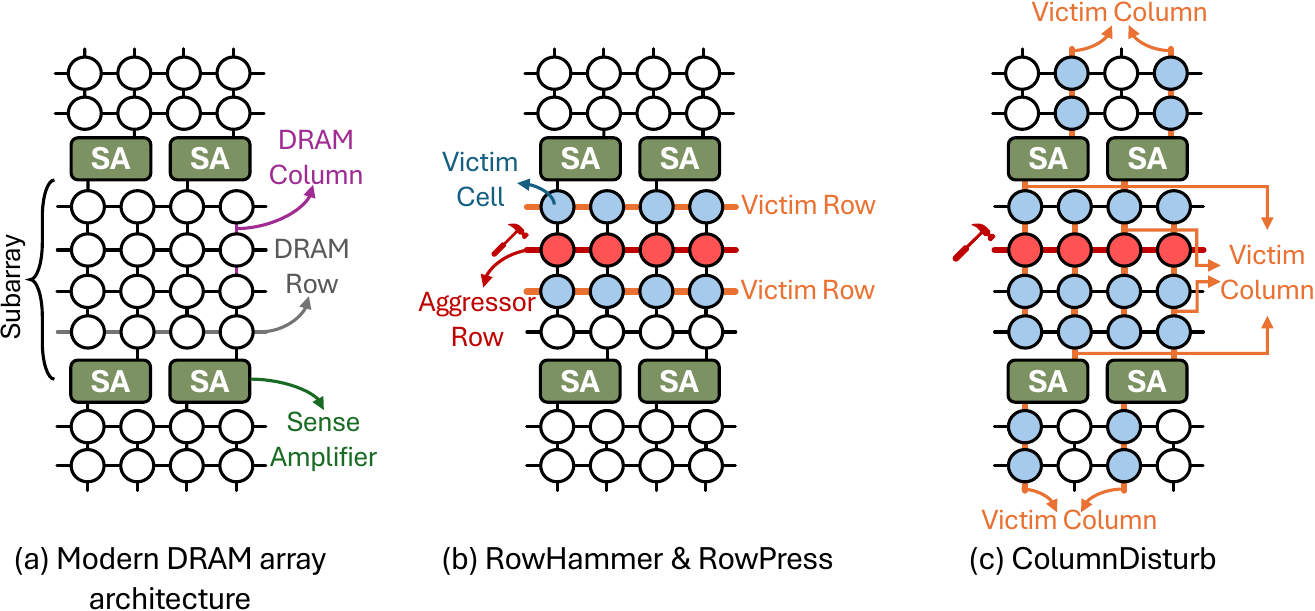}
    \caption{\ieycr{2}{(a)} Modern DRAM array architecture, \ieycr{2}{(b)} RowHammer~\&~RowPress induce bitflips in a few neighboring rows of the aggressor row\ieycr{2}{, where rows are the victims,} \ieycr{2}{(c)} \X{}, a column-based read disturbance phenomenon that affects many rows in multiple subarrays\ieycr{2}{, where columns are victims.}}
    \label{fig:mech_intro}
\end{figure}

\ieycr{1}{\figref{fig:mech_intro} highlights the \omcr{2}{conceptual} difference between \X{} (\figref{fig:mech_intro}c) and RowHammer~\&~RowPress (\figref{fig:mech_intro}b) in modern DRAM chips (\figref{fig:mech_intro}a).}

\ieycr{2}{RowHammer~\&~RowPress affect \emph{a few} rows within the same subarray (e.g., two victim rows in \figref{fig:mech_intro}b). RowHammer~\&~RowPress happen because rows are too close to each other, affecting cells that are in closeby rows (i.e., closeby rows are victims). In contrast, \X{} is fundamentally different: \X{} affects thousands of rows, including distant ones, across both the same subarray and multiple neighboring subarrays (e.g., \figref{fig:mech_intro}c). This is because accessing a DRAM row (i.e., DRAM row activation) perturbs many columns, and columns of a single DRAM row span multiple subarrays. Thus, \X{} affects cells that are vertically connected to the same bitline (i.e., columns are victims).}

\ieycr{2}{Our experimental results (\secref{sec:foundational} and \secref{sec:indepth}) and 
hypothetical explanations (\secref{sec:avg_bl_voltage}) for \X{} strongly suggest that \X{} induces bitflips
due to \emph{bitline-voltage-induced disturbance}. In other words, ``hammering'' (repeatedly raising and lowering) or ``pressing'' 
(keeping high or keeping low) a \emph{bitline's voltage} is
what leads to \X{} bitflips.}

\begin{figure}[ht]
    \centering
    \includegraphics[width=0.96\linewidth]{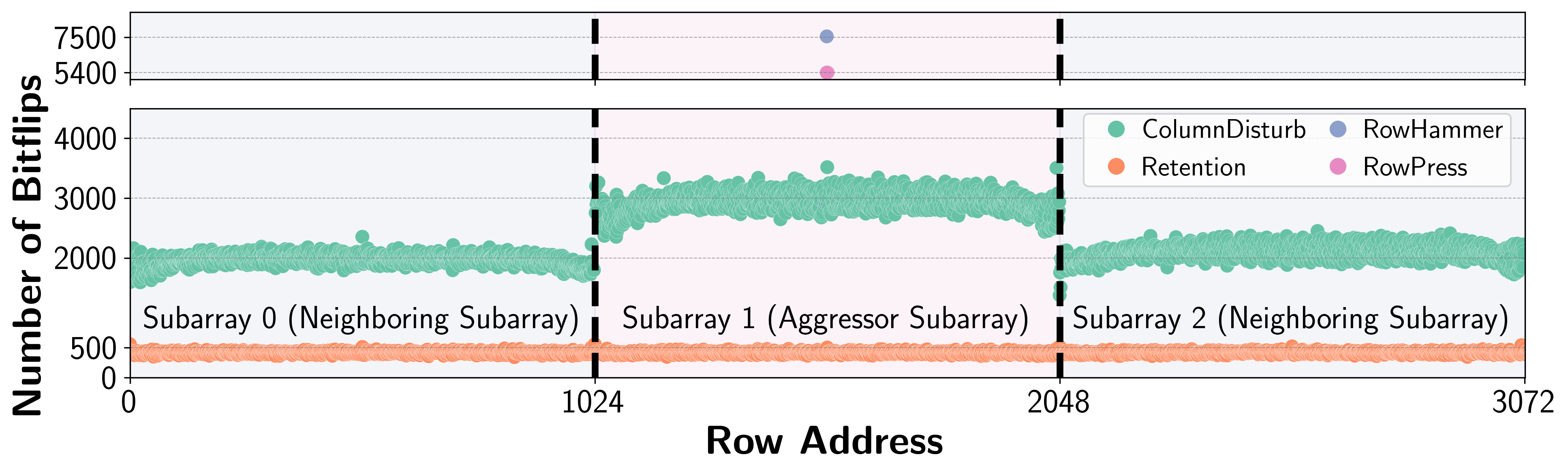}
    \caption{\ieycr{1}{Number of \X{}, RowHammer, RowPress, and retention failure bitflips in  three subarrays.}}
    \label{fig:bitflips_extended}
\end{figure}

To \omcr{2}{empirically} illustrate the \X{} phenomenon, \figref{fig:bitflips_extended} shows the \agy{bitflips caused by} \X{}\ieycr{1}{, RowHammer, RowPress,} and by retention \ieycr{2}{failures} in the first 3072 rows of a \ieycr{0}{representative Samsung 16Gb A-die DDR4 DRAM} module, spanning the first three consecutive subarrays. \ieycr{2}{The x}-axis shows the DRAM row address\ieycr{2}{, and the }y-axis shows the number of \ieycr{1}{\X{}, RowHammer, RowPress, and retention failure bitflips in each row.}\footnote{\ieycr{2}{Please see \secref{sec:disturbing_columns} for the detailed methodology and observations.} We study \X{} in great detail in \secref{sec:foundational} and \secref{sec:indepth}.}

We observe that \X{} \ieycr{1}{induces bitflips across three consecutive subarrays, and these bitflips are \omcr{3}{\emph{not} due to} retention failure\omcr{3}{s}. \omcr{2}{Nor are they \omcr{3}{due to} RowHammer~\&~RowPress, which affect only} immediate neighbor\ieycr{2}{ing} rows of the aggressor row.}

Based on our rigorous \omcr{1}{experimental} characterization of \X{} and its behavior under numerous parameters (\ieycr{1}{i.e., temperature, data pattern, DRAM timing parameters, average voltage level of the bitline, memory access pattern, and spatial variation}), we make \param{27} empirical observations and share \param{12} key takeaway lessons. We highlight \param{two} of our major new results.

First, newer DRAM chips manufactured with smaller technology nodes are increasingly \ieycr{0}{more} vulnerable to \X{} across all three major manufacturers. For example, we find that the minimum
time to induce \ieycr{0}{the} first \X{} bitflip reduces \omcr{0}{by} up to 5.06x \ieycr{1}{and \param{2.96}x on average across all tested modules}. We observe that even in \emph{existing} \gls{cots} DRAM chips\ieycr{2}{,} \X{} induces bitflips within a nominal DDR4 refresh {window} (e.g., in 63.6ms) in multiple cells from a single module. Second, significantly more rows experience \X{} bitflips than retention failures, and the number of \X{} bitflips is much higher than retention failures in all tested \ieycr{0}{refresh intervals (i.e., 64ms, 128ms, 256ms, 512ms, and 1024ms)} across all tested DRAM modules. For example, \omcr{0}{with a} 512ms \omcr{0}{refresh interval} and at \SI{65}{\celsius}, \X{} induces \ieycr{1}{1.64x, 62.49x, and} \param{152.66}x more bitflips in \ieycr{1}{2, 6, and} \param{232} more rows on average for \atb{tested} \ieycr{1}{SK Hynix, Micron, and} Samsung modules\ieycr{1}{, respectively,} \omcr{0}{than retention failures}. 

Our observations show that \X{} \ieycr{0}{has} serious implications \omcr{0}{on} the robustness of \omcr{0}{both} 1) future systems and 2) existing retention-aware heterogeneous refresh mechanisms~\omcr{1}{\retentionMechanisms{}}. 
First, due to continuously shrinking DRAM node size, more \X{} bitflips \omcr{0}{may} manifest in the standard refresh {window} \omcr{0}{in future DRAM chips}, jeopardiz\omcr{0}{ing} the robustness of future systems. We \atb{\omcr{0}{describe and evaluate} \param{two}} \omcr{0}{hardware} techniques that could mitigate \X{} bitflips at varying expected performance, energy, and area overheads. \atb{A straightforward solution \omcr{0}{is to} {increase \omcr{1}{the} DRAM refresh rate}
to accommodate \X{} bitflips that could happen in the standard refresh {window}\ieycr{0}{. However, this straightforward solution} \omcr{1}{reduces system throughput by 42.1\%} and \ieycr{1}{increases system energy consumption by} \param{67.5\%}. We \omcr{0}{instead} propose
to intelligently and timely refresh \omcr{0}{\emph{only the victim rows}} that are
\omcr{0}{vulnerable to} \X{}\ieycr{1}{. We} show that \ieycr{1}{our} \omcr{0}{improved} solution \ieycr{1}{reduces the straightforward solution's 1) system throughput overhead by} \param{\atbcr{3}{70.5}\%} and 2) energy overhead by \param{73.8\%}.}
Second, we evaluate a retention-aware refresh mechanism \omcr{0}{(RAIDR)}~\cite{liu2012raidr} and demonstrate that its benefits drastically decrease in the presence of \X{} (e.g., \param{53}\% decrease in performance)
\ieycr{0}{compared to a baseline RAIDR that does not suffer from \X{}.}

\atbcr{0}{We call for future research to \ieycr{2}{1) fundamentally understand \X{} at the device-level, 2)} architect \X{}-resilient, high-performance retention-aware refresh mechanisms\ieycr{1}{, and\ieycr{2}{ 3)} other innovative solutions to mitigate \X{} bitflips to enable \X{}-resilient, robust future computing systems.}}

This paper makes the following key contributions:
\begin{itemize}
    \item \omcr{1}{T}his is the first work to experimentally demonstrate a \emph{column-based} \omcr{1}{(i.e.,~bitline-based)} read disturbance phenomenon \omcr{0}{in modern DRAM chips}, \X{}, and its widespread existence in real DDR4 \omcr{1}{chips} \ieycr{0}{from all three major \omcr{2}{DRAM} manufacturers and HBM2 chips from Samsung}. 
    \item We provide an extensive \omcr{0}{experimental} characterization of \X{} on \nCHIPS{} real \ieycr{0}{DDR4} and 4 HBM2 DRAM chips. \ieycr{1}{\X{} induces bitflips across three subarrays (e.g., 3072 rows), \omcr{2}{greatly} more than RowHammer~\&~RowPress that affect \omcr{2}{only} a few rows \ieycr{2}{in a single subarray}.} 
    \item Our \omcr{2}{experimental} results show that \X{} 1) gets worse as DRAM technology scales down \omcr{1}{to} \omcr{0}{smaller cell sizes, 2)} \omcr{1}{already} induces bitflips within a standard refresh {window} \omcr{0}{in some existing DRAM chips}\ieycr{2}{,} and \omcr{0}{3)} induces significantly more bitflips in many more rows than retention failures\ieycr{0}{, for a given refresh interval}.
    \item We \omcr{1}{describe} and evaluate two solutions to \X{} for future DRAM-based systems.
    \item \omcr{0}{We} evaluate a retention-aware heterogeneous refresh mechanism \ieycr{3}{(RAIDR~\cite{liu2012raidr})} and show that \X{} can completely diminish the performance and 
    energy benefits of such mechanisms.
\end{itemize}

\glsresetall
\section{Background \omcr{1}{\& Fundamentals}}

\subsection{Dynamic Random Access Memory (DRAM)}
\label{sec:dram_organization}

\begin{figure}[ht]
    \centering
    \includegraphics[width=\linewidth]{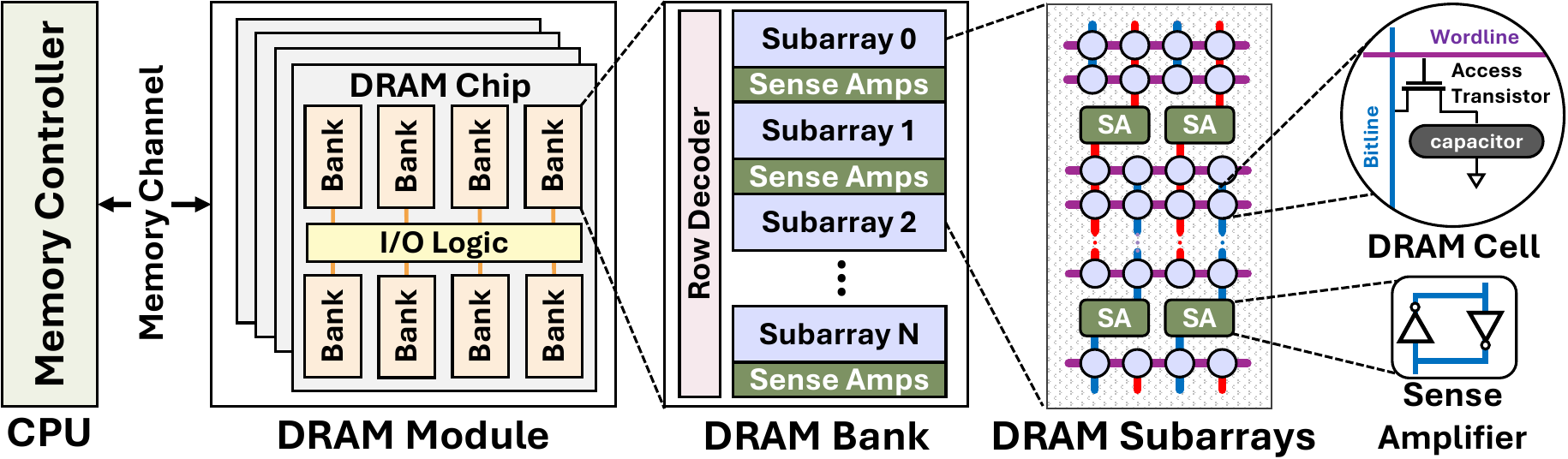}
    \caption{{\omcr{0}{O}rganization of modern DRAM\omcr{0}{-based main memory}.}}
    \label{fig:dram_organization}
\end{figure}

\noindent\textbf{DRAM Organization.}~\figref{fig:dram_organization} shows the hierarchical organization of modern DRAM-based main memory. A module contains one or multiple DRAM ranks that time-share the memory channel. A rank consists of multiple DRAM chips.
Each DRAM chip has multiple DRAM banks, each containing multiple subarrays~\cite{chang2016low,seshadri2013rowclone,kim2012case}. 
Within a subarray, DRAM cells form a two-dimensional structure interconnected over \textit{bitlines} and \textit{wordlines}. Each cell encodes one bit of data using the charge level in its capacitor. A true cell encodes data ‘1’ as fully charged (VDD) and data ‘0’ as fully discharged (GND), whereas an anti-cell uses the opposite encoding. The data in the cell is accessed through an access transistor, driven by the wordline to connect the cell capacitor to the bitline. The row decoder decodes the row address and drives one wordline. A row of DRAM cells on the same wordline is referred to as a DRAM \emph{row}. DRAM cells in the same \emph{column} are connected to the sense amplifier via a bitline.
To sense \omcr{0}{an} entire row of cells, each subarray has bitlines that connect to two rows of sense amplifiers, one above and one below the subarray, which causes neighboring subarrays to share half of the bitlines~\cite{dram-circuit-design,schloesser20086f,sekiguchi2002low,luo2020clrdram,chang2016lisa,jacob_book_2008,itoh2013vlsi,itoh77, lee2013tiered}. As a result,
{a subarray’s even bitlines (highlighted in red for the middle subarray\ieycr{0}{, Subarray 1 in \figref{fig:dram_organization}}) are connected to one of its neighboring subarrays’ odd bitlines (highlighted in red for the top subarray\ieycr{0}{, Subarray 0 in \figref{fig:dram_organization}}), while the odd bitlines of Subarray 1 (highlighted in blue for Subarray 1) are connected to the other neighboring subarray’s even bitlines (highlighted in blue for the bottom subarray\ieycr{0}{, Subarray 2 in \figref{fig:dram_organization}}) via sense amplifiers.} This approach, known as the {\emph{open-bitline architecture}}, is widely adopted {in} high-density DRAM~\cite{dram-circuit-design,schloesser20086f,sekiguchi2002low,luo2020clrdram,chang2016lisa,jacob_book_2008,itoh2013vlsi,itoh77, lee2013tiered}.

\noindent\textbf{DRAM Access.}
Accessing DRAM consists of two steps. First, the memory controller issues an \act{} (activate) command together with a row address to the bank. 
The row decoder drives the wordline of that row to activate the row (i.e., enables the access transistors). 
Data is then transferred from the DRAM cells in the row to the row buffer through the bitlines. 
Second, the memory controller sends a \pre{} (precharge) command to close the activated row after waiting for $t_{RAS}$ (i.e., the minimum time between opening a row with an \act{} command and closing the row with a \pre{} command). Before accessing another row in the same bank, the memory controller must wait for $t_{RP}$ (i.e., the minimum time between sending a \pre{} command and opening a row with an \act{} command).

\subsection{DRAM Read Disturbance}
Read disturbance is the phenomenon that reading data from a memory or storage device causes physical disturbance (e.g., voltage deviation, electron injection, electron trapping) on another piece of data that is \emph{not} accessed but located \omcr{0}{physically} nearby the accessed data. Two prime examples of read disturbance in modern DRAM chips are RowHammer~\cite{kim2014flipping,mutlu2017rowhammer,mutlu2019rowhammer,mutlu2023fundamentally,luo2023rowpress,mutlu2023retrospective} and RowPress~\cite{luo2023rowpress,luo2024rowpress}, where repeatedly accessing (hammering) or keeping active (pressing) a row induces bitflips in physically nearby rows, respectively. For read disturbance bitflips to occur, 1)~the aggressor row needs to be activated more \ieycr{0}{times} than a certain threshold value~\cite{kim2014flipping}, and/or 2) the aggressor row needs to be open for a long period of time  (i.e., $\DRAMTIMING{AggOn} > \DRAMTIMING{RAS}$){~\cite{luo2023rowpress}}.

\subsection{DRAM Refresh and Retention Failures}
\label{subsec:back_retention}
\noindent\textbf{DRAM Refresh.}
DRAM cell capacitors inherently lose charge over time, potentially resulting in
data corruption\ieycr{0}{~\cite{liu2012raidr,qureshi2015avatar,liu2013experimental}}. A DRAM cell’s retention time defines how long it can
reliably store data and typically varies between cells from milliseconds to many hours~\cite{liu2012raidr,qureshi2015avatar,liu2013experimental,patel2017reach,khan2014efficacy,weis2015retention,jung2017platform,khan2017detecting}. To prevent retention \ieycr{0}{failures} from appearing, the memory controller periodically restores each DRAM row’s charge level by sending \omcr{0}{a} \texttt{REF} (refresh) command every $t_{REFI}$\ieycr{0}{, e.g.,~\cite{jedec2012ddr3, jedec2020ddr4, micron2014ddr4, hassan2021utrr, liu2013experimental}}, which \omcr{0}{is} scheduled according to a timing parameter called the refresh window ($t_{REFW}$)~\ieycr{0}{~\cite{jedec2012ddr3, jedec2020ddr4,
micron2014ddr4,liu2013experimental}}.

\noindent
\hluo{\noindent\textbf{DRAM Retention Failures.} \omcr{0}{A m}odern DRAM cell with Buried Channel Access Transistor (BCAT)~\cite{Yang2013SuperiorImprovements, Park2015Technology} has multiple leakage paths~\cite{Liu2024UnderstandingRetention, Roy2003Leakage, lee2011simultaneous}. Prior works show that the major leakage paths include Gate-Induced Drain Leakage and Junction Leakage~\cite{saino00, Yang2013SuperiorImprovements, Park2015Technology, lee2011simultaneous,yaney1987meta,restle1992dram}. These major leakage paths are all from the cell to the substrate of the access transistor, which is biased at a negative voltage to reduce subthreshold leakage through the transistor~\cite{lee2011simultaneous}, causing true-cells storing data `1' much more likely to experience retention failures than `0'~\cite{liu2013experimental,patel2017reaper,qureshi2015avatar}. As DRAM \omcr{0}{technology} node \omcr{0}{size reduces}, the close\omcr{0}{r} proximity between the capacitor contact and the bitline also creates potential leakage paths~\cite{Yu2022TheStudy}.} 

\section{\omcr{0}{Evaluation Infrastructure \&} Methodology}
\omcr{2}{W}e describe our \gls{cots} DRAM testing infrastructure (\secref{subsec:infra}) and our testing methodology (\secref{subsec:method}). 

\subsection{COTS DRAM Testing Infrastructure}
\label{subsec:infra}
\begin{figure}[ht]
\centering
\includegraphics[width=\linewidth]{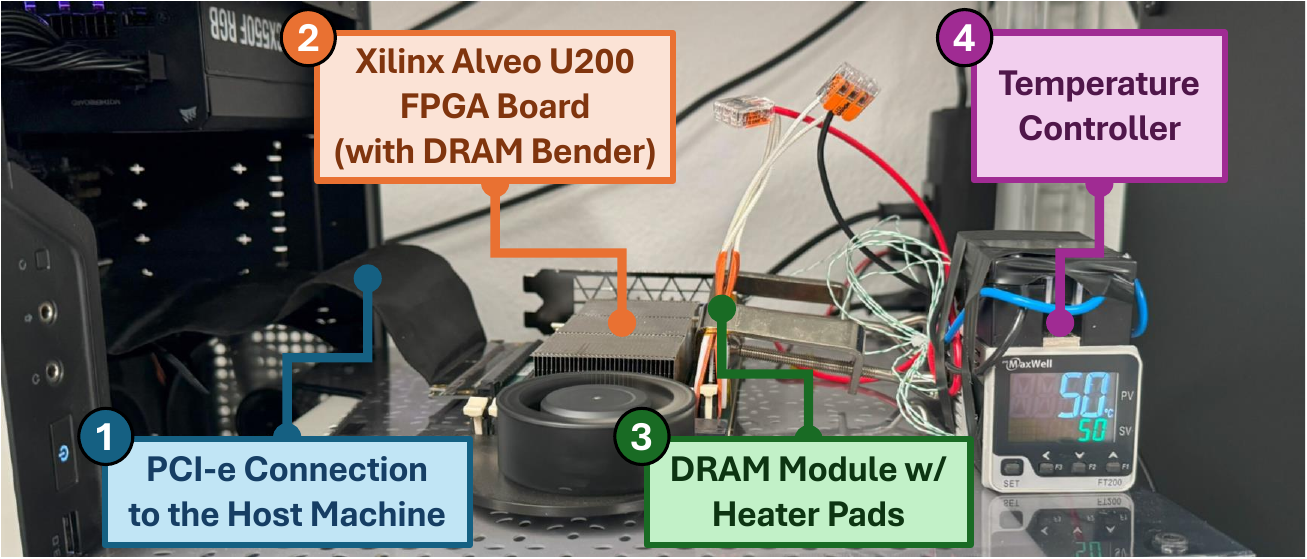}
\caption{\ieycr{0}{Our DRAM Bender~\cite{olgun2023dram} based experimental setup.}}
\label{fig:infra}
\end{figure}

\begin{figure}[ht]
\centering
\includegraphics[width=\linewidth]{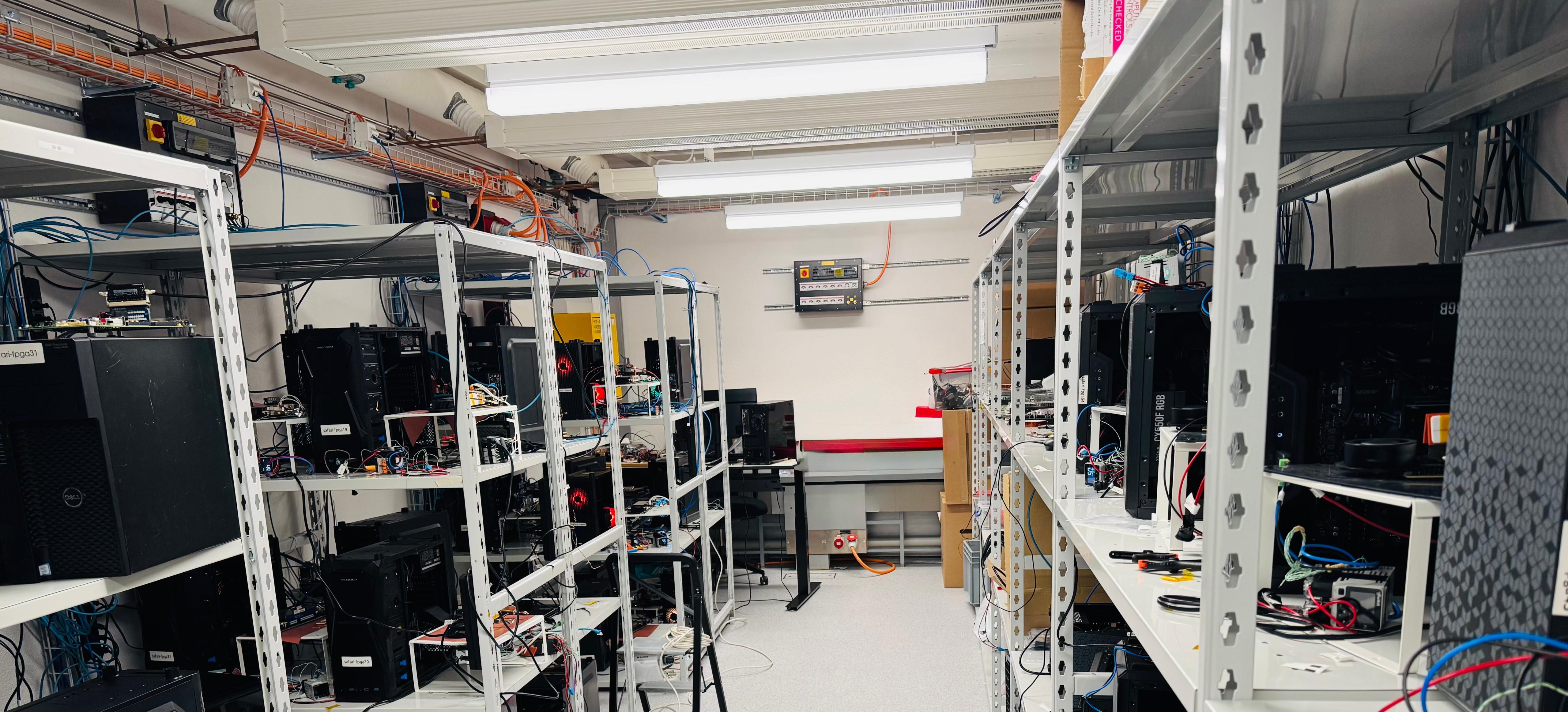}
\caption{\ieycr{0}{Our laboratory \omcr{2}{for} real DRAM chip experiments.}}
\label{fig:lab}
\end{figure}

We conduct COTS DRAM chip experiments using DRAM Bender~\cite{safari-drambender, olgun2023dram} \omcr{0}{(built upon SoftMC~\cite{hassan2017softmc,softmcgithub})}, an FPGA-based DDR4 and HBM2 testing infrastructure that provides precise control of DDR4 and HBM2 commands issued to a DRAM module. \ieycr{0}{\figref{fig:infra} shows one of our experimental setups on DDR4 modules} that consists of four main components: 1) a host machine that generates the test program and collects experiment results, 2) an FPGA development board~\cite{alveo_u200}, 3) a temperature sensor and heater pads pressed against the DRAM chips to maintain a target temperature level, and 4) a temperature controller~\cite{maxwellFT200} to control the temperature. \ieycr{0}{\figref{fig:lab} shows our laboratory \omcr{1}{comprising} many DDR4/HBM2 testing platforms.}

\noindent\textbf{Real DDR4 and HBM2 DRAM Chips Tested.}
\label{sec:chips}
Table~\ref{tab:dram_chips} provides \nCHIPS{} real DDR4 chips from \nMODULES{} modules and 4 HBM2 chips \omcr{0}{that we characterize}.\footnote{\label{fn:node}The technology node \omcr{0}{size} of a DRAM chip is usually not publicly available. Prior works~\cite{luo2023rowpress,kim2020revisiting,orosa2021deeper,yuksel2025pudhammer,olgun2025variable,tugrul2025understanding} assume that for a given chip manufacturer and chip
density, the alphabetical order of die revision codes may provide an indication
of technology node advancement.}\textsuperscript{,}\footnote{\ieycr{1}{We provide much more detail on the tested DRAM chips in the extended version of this paper~\cite{yuksel2025columndisturb}.}}

%\begin{scriptsize}
%216
\begin{table}[ht]
\centering
\caption{Summary of DDR4 and HBM2 DRAM chips tested.}
\label{tab:dram_chips}
\resizebox{\linewidth}{!}{%
\renewcommand{\arraystretch}{0.87}
\begin{threeparttable}
\begin{tabular}{cccccc}
\textbf{Chip Mfr.} & \textbf{Module IDs} & \textbf{\#Chips} &  \textbf{Die Rev.} & \textbf{Density} & \textbf{Org.}  \\
%\textbf{Mfr.} & \textbf{(\#Chips)} & \textbf{Rev.} & \textbf{Dens.} & \textbf{Org.}  & \textbf{Size}          %\\ 
\hline\hline
         & H0,H1,H2  & \param{24} & A & 8Gb  & x8 \\
SK Hynix & H3,H4,H5,H6  & \param{32} & D & 8Gb  & x8  \\ 
         & H7  & \param{8} & A & 16Gb  & x8   \\
         & H8,H9 & \param{16} & C & 16Gb  & x8 \\

\midrule

        & M0  & \param{8} & B  & 4Gb & x8 \\
        & M1,M2,M3  & \param{24} & R & 8Gb & x8  \\
Micron  & M4, M5  & \param{16} & B  & 16Gb & x8 \\
        & M6, M7  & \param{8} & E  & 16Gb & x16 \\
        & M8,M9,M10,M11  & \param{32} & F  & 16Gb & x8 \\

\midrule
            & S0, S1  & \param{16} & A  & 16Gb & x8 \\
Samsung     & S2, S3  & \param{16} & B  & 16Gb & x8 \\
            & S4, S5  & \param{16} & C  & 16Gb & x16 \\

\midrule
\midrule

Samsung     & HBM2 Chips  & \param{4} & N/A  & N/A & N/A \\
        
\bottomrule
\end{tabular}
\begin{tablenotes}
\item[a] We report “N/A” if the information is not publicly available.
\end{tablenotes}
\end{threeparttable}
}
\end{table}

\noindent\textbf{Logical-to-Physical Row Mapping.}
DRAM manufacturers use mapping schemes to translate logical
addresses to physical row addresses~\cite{kim2014flipping, smith1981laser, horiguchi1997redundancy, keeth2001dram, itoh2013vlsi, liu2013experimental,seshadri2015gather, khan2016parbor, khan2017detecting, lee2017design, tatar2018defeating, barenghi2018software, cojocar2020rowhammer,  patel2020beer}. To account for in-DRAM row address mapping {and internal row remapping (which are extensively discussed in~\cite{hassan2021utrr,nam2024dramscope,kim2020revisiting})}, we reverse engineer the physical row address layout, following prior works' methodology~\cite{kim2020revisiting, orosa2021deeper, yaglikci2022understanding, luo2023rowpress, hassan2021utrr, yaglikci2022hira, yaglikci2024svard, nam2024dramscope}.

\subsection{Experimental Methodology}
\label{subsec:method}
Characterizing and understanding \X{} requires 1) reverse engineering the subarray boundaries \ieycr{0}{because \ieycr{1}{the columns of a single DRAM row span three physically consecutive subarrays, and thus,} \X{}} bitflips can manifest \ieycr{0}{across} three consecutive subarrays \ieycr{1}{(as shown in \figref{fig:bitflips_extended})}, 2) filtering \ieycr{0}{out} retention failures\ieycr{0}{, since} \X{} induces bitflips \omcr{1}{within intervals} longer than \nb{the} refresh window ($\trefw{}$), and 3) testing many parameters to understand how operational conditions and parameters affect the \X{} vulnerability.

\noindent\textbf{\nb{Metrics}.}
To characterize a DRAM module's vulnerability to \X{}, we examine three metrics: \one{} minimum time to induce the first \omcr{0}{\X{}} bitflip in a subarray, \two{} the fraction of cells with bitflips in a subarray, and \three{} the number of rows that experience at least one \omcr{0}{\X{}} bitflip in a subarray (i.e., blast radius~\cite{kim2014flipping,kim2020revisiting,orosa2021deeper,yaglikci2022understanding,luo2023rowpress, hassan2021utrr, yaglikci2022hira, yaglikci2024svard, olgun2025variable, tugrul2025understanding,lang2023blaster,kogler2022half}). A \omcr{0}{shorter} time to induce the first bitflip indicates \omcr{0}{higher} vulnerability to \X{}. \omcr{0}{H}igher values for the fraction of cells with bitflips in a subarray and blast radius indicate higher \omcr{0}{\X{}} vulnerability.

\noindent\textbf{Access Pattern.} To induce \X{} bitflips in a subarray, we perform the following key DRAM command sequence: 

\act{} $R_{Agg}$ $\xrightarrow[]{t_{Agg_{On}}}$ \pre{} $\xrightarrow[]{t_{RP}}$ \act{} $R_{Agg}$ $\xrightarrow[]{t_{Agg_{On}}}$ $\cdots$

\noindent where $R_{Agg}$ is the aggressor row in the tested subarray, $t_{Agg_{On}}$ is \nb{the duration} that \omcr{1}{the} aggressor row stays open (i.e., the timing delay between \act{} $R_{Agg}$ and \pre{})~\cite{luo2023rowpress}, and $\trp{}$ is the timing delay before issuing the next \act{} $R_{Agg}$ command. \ieycr{0}{We conduct all experiments where $R_{Agg}$ is the middle row in the tested subarray unless stated otherwise.} 

\ieycr{0}{By repeatedly hammering or pressing $R_{Agg}$ with our access pattern \ieycr{1}{(e.g., the aggressor row in \figref{fig:mech_intro}-c)}, we disturb DRAM cells through the columns \ieycr{1}{(e.g., victim columns in \figref{fig:mech_intro}c)}, since a row drives its content to all columns in a subarray after a successful ACT command. As a result, hammering repeatedly raises and lowers the voltage of the columns, while pressing keeps the voltage consistently high or low.}
\omcr{0}{We conduct a}ll experiments using the same access pattern \ieycr{2}{and record the bitflips in the aggressor row's subarray} unless stated otherwise.

\noindent\textbf{\omcr{1}{Algorithm to Find} Time to Induce \omcr{2}{the} First Bitflip.} For every 
parameter we evaluate, we 
\nb{determine} the minimum \ieycr{0}{number of ACT commands (i.e., hammer count) required to} induce the first bitflip 
\nb{per} tested
subarray\nb{,} using the bisection-method algorithm used by prior works~\cite{orosa2021deeper,yaglikci2022understanding,luo2023rowpress,tugrul2025understanding,yuksel2025pudhammer,yaglikci2024svard,luo2024rowpress}. We terminate the search when the difference between the current and previous minimum hammer count measurements is \nb{smaller} than 1\% of the previous measurement. \ieycr{1}{During these experiments, we do not issue any REF commands for 512 ms (i.e., the refresh interval is 512 ms). If no \X{} bitflips are observed within 512 ms in a given subarray, we terminate the search for that subarray and proceed to the next one}. For every tested subarray, we repeat the search five times and convert the minimum hammer count value across five iterations to time.

\noindent\textbf{DRAM Subarray Boundaries.} 
Prior \agy{works}~\cite{gao2019computedram,olgun2022pidram,olgun2021quac,yuksel2024functionally,yuksel2024simultaneous,yaglikci2024svard,mutlu2024memory,mutlu2025memory} \agy{demonstrate} that real DRAM chips are capable of performing the RowClone operation~\cite{seshadri2013rowclone}, \ieycr{0}{an in-DRAM data copy operation within a subarray by performing two consecutive row activations, which leads to data in the first activated row (i.e., source row) to be copied to the second activated row (i.e., destination row) via the sense amplifiers connected to both rows. We repeatedly perform the RowClone operation for \emph{every} possible source and destination row address in each tested bank. When we observe that the destination row gets the same content as the source row after the RowClone operation, we conclude that the source row and the destination row are in the same subarray.}
Based on this observation, we reverse engineer the subarray boundaries and determine which rows are in the same subarray.

\noindent\textbf{Filtering \omcr{0}{Out} Retention and RowHammer \& RowPress Failures.} 
\agy{We experimentally identify cells that experience retention failures during our tests and exclude them from the set of bitflips caused by \X{} in two steps.}
First, we \agy{follow the state-of-the-art retention time test methodology~\cite{patel2017reaper,qureshi2015avatar,liu2013experimental,khan2014efficacy} for \param{five} different data patterns.}
Second, \agy{to cover the worst-case in the presence of the variable retention time phenomenon\omcr{0}{~\cite{restle1992dram, qureshi2015avatar,yaney1987meta,liu2013experimental}}, we repeat each test 50 times and draw our conclusions based on the \omcr{0}{lowest observed} retention time of a cell.}
To filter \omcr{0}{out} RowHammer and RowPress bitflips, in every experiment, we exclude eight nearest physical victim rows of the aggressor row when we read from a subarray, \ieycr{1}{since RowHammer~\&~RowPress induce bitflips in \omcr{2}{nearby} neighbor\omcr{2}{ing} rows of the aggressor row. \ieycr{2}{We exclude eight nearest neighbor rows of the aggressor, even though we \omcr{3}{experimentally find that} RowHammer~\&~RowPress induce bitflips in only +1/-1 neighboring rows. We add this guardband (i.e., six \omcr{3}{additional} rows to exclude) because the state-of-the-art industry solutions to DRAM read disturbance \ieycr{3}{support} refresh\ieycr{3}{ing} RowHammer \& RowPress bitflips in up to eight nearest neighboring rows~\cite{jedec2020ddr5,jedec2024jesd795c,hassan2021utrr,frigo2020trrespass,canpolat2024understanding}.\footnote{\ieycr{3}{We also observe that when we only exclude two nearest neighbor rows, our experimental results are \omcr{4}{essentially} the same. Our extended version provides more comprehensive analyses and results~\cite{yuksel2025columndisturb}}.}}}

\noindent\textbf{Eliminating Other Interference Sources.}
We eliminate other potential sources of interference by taking \param{two} measures, similar to prior works~\cite{kim2020revisiting, orosa2021deeper, yaglikci2022understanding, hassan2021utrr, luo2023rowpress}.
First, we disable periodic refresh to prevent refreshing any rows so that we can observe the DRAM \omcr{0}{inherent device-level} behavior.
Second, we {verify} that the tested modules and chips have neither rank-level nor on-die ECC~\cite{patel2020beer, patel2021harp,mineshphd}.

\noindent\textbf{Test Parameters.} We explain the three common parameters \omcr{0}{we use in testing} and elaborate on the 
\nb{specific} parameters of each test in \secref{sec:foundational} and \secref{sec:indepth}.
\noindent\textit{1)~Data Pattern:}
\nb{We} use five data patterns 0x00, 0xAA, 0x11, 0x33, and 0x77\omcr{0}{, commonly} used in memory reliability testing~\cite{vandegoor2002address} and by prior work on DRAM characterization (e.g.,~\cite{kim2014flipping,kim2020revisiting,orosa2021deeper,luo2023rowpress,yaglikci2024svard}). We fill aggressor rows with these data patterns while initializing victim rows with the negated data pattern (e.g., if aggressors have~{0x00}, victim rows are filled with~{0xFF}).
\noindent\textit{2) Temperature:} 
We perform our experiments at four temperature levels: 45$^{\circ}$C, 65$^{\circ}$C, 85$^{\circ}$C, and 95$^{\circ}$C. \ieycr{0}{We conduct all} experiments at 85$^{\circ}$C unless stated otherwise.
\noindent\textit{3) Aggressor Row On Time (\DRAMTIMING{AggOn}):} \nb{We define \DRAMTIMING{AggOn}~\cite{luo2023rowpress} as the} time an aggressor row stays open after each activation during a \X{} test. We perform our experiments at four \DRAMTIMING{AggOn} values: 36$ns$, 7.8$\mu$$s$, 70.2$\mu$$s$, and 1$ms$. \ieycr{0}{We conduct all} experiments at \DRAMTIMING{AggOn}$=$70.2$\mu$$s$ unless stated otherwise. \ieycr{1}{\noindent\textit{4) Refresh Interval:} We perform our experiments with various different refresh intervals from 64ms to 16s.}

\setstretch{0.92}
\section{Foundational Results}
\label{sec:foundational}
We demonstrate a new read disturbance phenomenon\omcr{1}{, \X{},} in real DDR4 and HBM2 \omcr{0}{DRAM} chips. \ieycr{1}{\X{} is a widespread read disturbance phenomenon in real DRAM chips and worsens as DRAM technology scales down
to smaller node sizes (\secref{sec:time2first_bitflip}). We observe that} \X{} can induce bitflips in three consecutive subarrays by hammering or pressing an aggressor row (\secref{sec:disturbing_columns}).
We investigate the characteristics of \X{} by analyzing the direction of bitflips (\secref{sec:bitflip_direction}), the effect of data pattern on \hluo{a DRAM column} (\secref{sec:dp_column}), the effect of aggressor row on time (\secref{sec:tAggON}), the effect of average voltage level \ieycr{0}{on a DRAM column} (\secref{sec:avg_bl_voltage}), the number of DRAM rows that experience \X{} bitflips (\secref{sec:blast_radius}), and~\X{} on HBM2 DRAM chips (\secref{sec:hbm}).

\subsection{\omcr{1}{Prevalence \& Scal\omcr{3}{ing Characteristics}}}
\label{sec:time2first_bitflip}

We study the vulnerability of each chip to \X{}. One critical component of vulnerability is identifying the weakest cell (i.e., the cell that fails \omcr{1}{earliest}). Since \X{} is effective \omcr{0}{at} subarray granularity \ieycr{1}{(see \secref{sec:intro} and \secref{sec:disturbing_columns})}, we check the subarray that the aggressor row \ieycr{0}{(the middle row in the tested subarray\footnote{\ieycr{1}{We test different aggressor row locations and observe the almost same trend in \secref{sec:id_loc}.}})} belongs to and record the time to induce the first \X{} bitflip in that subarray for each chip \ieycr{1}{(see \secref{subsec:method} for more details on \omcr{3}{our complete} methodology)}. \ieycr{0}{In this experiment, we test \omcr{2}{\emph{all}} subarrays in \omcr{2}{\emph{all}} banks across \omcr{2}{\emph{all}} tested DRAM modules from \omcr{2}{\emph{all}} three major manufacturers \ieycr{2}{using the parameters with which a DRAM chip experiences the highest ColumnDisturb vulnerability (determined through extensive experiments in \secref{sec:foundational} and \secref{sec:indepth}).}}

\begin{figure}[ht]
    \centering
    \includegraphics[width=\linewidth]{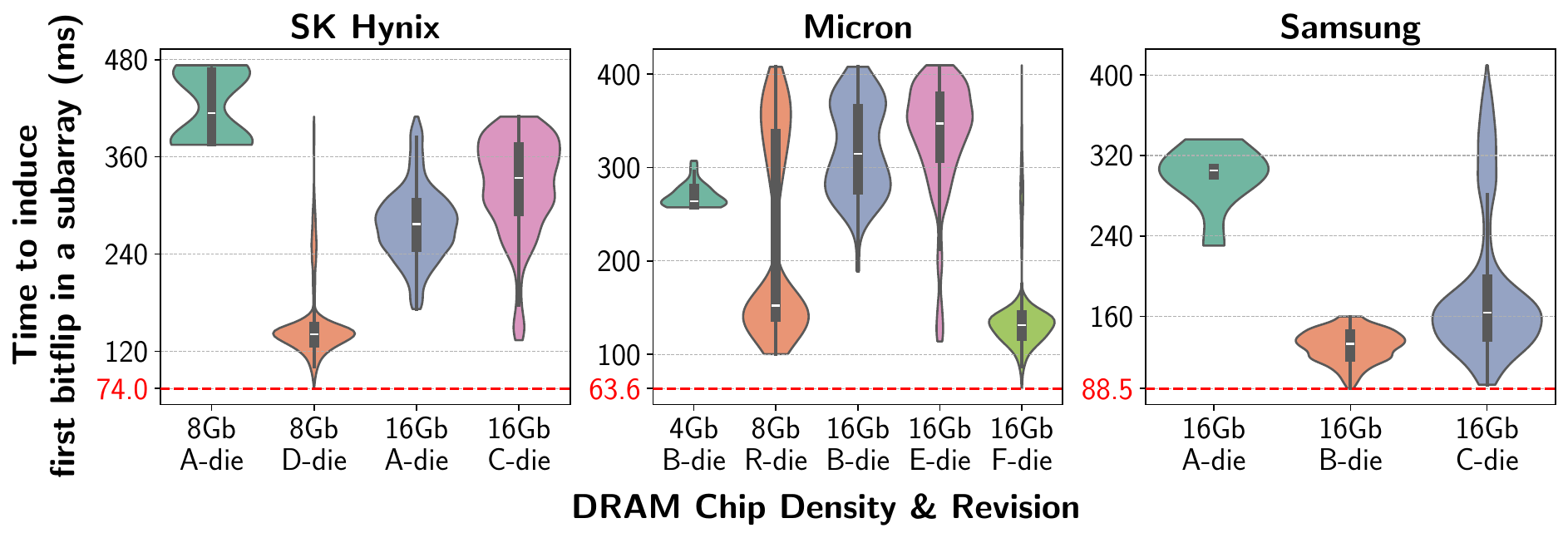}
    \caption{\ieycr{1}{Distribution of the time to induce the first \X{} bitflip in a subarray for different DRAM chip densities \& die revisions across three major manufacturers}.}
    \label{fig:hcfirst}
\end{figure}

\figref{fig:hcfirst} \ieycr{1}{shows a violin plot of} \omcr{0}{the} time to observe the first \ieycr{1}{\X{}} bitflip in a subarray \ieycr{0}{across all tested \ieycr{3}{46,080} subarrays} (y-axis) for each different die revision and density pair (x-axis).
Each subplot is dedicated to a different manufacturer.

\ieycr{1}{\observation{All tested \nCHIPS{} DDR4 chips are vulnerable to \X{}.}}
\ieycr{1}{We observe that in every DDR4 chip tested, there is at least one DRAM cell that experiences a \X{} bitflip.}

\observation{\label{obs:hcfirst_2} Newer chips tend to be more vulnerable to \X{} bitflips.}
We observe that in the same die density for each tested manufacturer, more advanced technology nodes experience \ieycr{1}{higher \X{} vulnerability (i.e.,} lower time to induce the first \X{} bitflip\ieycr{1}{)}.\footnote{For a given manufacturer and die density, the later in the alphabetical order the die revision code is, the more likely the chip has a more advanced technology node~\cite{luo2023rowpress,kim2020revisiting,orosa2021deeper}.} 
\omcr{1}{For} SK Hynix, for 8Gb and 16Gb chip density, respectively, the minimum time to induce the first \X{} bitflip in a subarray reduces by \param{5.06}\omcr{1}{x} and \param{1.29}x as die generation advances (i.e., 8Gb A- to D-die and 16Gb A- to C-die).
\omcr{1}{For} Micron, for 16Gb chip density, from B- to F-die, \ieycr{1}{the minimum time to induce the first \X{} bitflip in a subarray} reduces by \param{2.98}x. 
\omcr{1}{For} Samsung, compared to 16Gb \ieycr{0}{A}-die, the minimum time to induce the first \X{} bitflip in a subarray is \param{2.50}x \ieycr{0}{lower} in 16Gb \ieycr{0}{C}-die.

\ieycr{3}{We hypothesize that as DRAM technology node advances, the distance between the DRAM cell capacitor contact and the bitline decreases, strengthening tunneling effects between them~\cite{Yu2022TheStudy}.} \ieycr{0}{Thus, DRAM chips become more vulnerable to \X{} as DRAM technology node \omcr{1}{size reduces}. \ieycr{2}{\secref{sec:avg_bl_voltage} provides detailed analyses and hypotheses on the device-level mechanisms of \X{}.}}

\ieycr{0}{Observation 2} has serious implications for the future\ieycr{1}{ as} DRAM technology node sizes will continue to reduce and the time to induce the first \X{} bitflip will likely get smaller. \omcr{1}{\secref{sec:implications-for-future-systems}} discuss\omcr{1}{es} the implications of this observation.

\observation{\label{obs:hcfirst_1} \omcr{1}{T}here \omcr{2}{already} are \omcr{2}{some} \omcr{1}{real DRAM} chips \omcr{1}{where \X{} introduces bitflips within the \omcr{3}{nominal} refresh window under \omcr{3}{nominal} operating conditions}.}
We observe that in a single 16Gb F-die Micron module, multiple cells \ieycr{0}{(at least four cells)} from multiple chips \ieycr{0}{(at least three chips)} experience \X{} bitflips at 63.6ms \ieycr{0}{and 85$^{\circ}$C}, within the nominal refresh window for DDR4 chips, $\trefw{}$. \ieycr{2}{In contrast,} retention \ieycr{1}{failures} require at least 512ms \ieycr{0}{(at 85$^{\circ}$C)} to manifest for the same module (see \figref{fig:blast}). We also observe that the closest \ieycr{2}{(farthest)} victim row that experiences \omcr{2}{\X{}} bitflips within $\trefw{}$ is 374 \ieycr{2}{(446)} rows away from the aggressor row \ieycr{3}{where both victim row and the aggressor row are in the same subarray}.
\takeaway{\ieycr{1}{\X{} is a widespread read disturbance phenomenon in real DRAM chips. \X{} becomes worse as DRAM technology scales down to smaller node sizes\omcr{1}{.~\X{}} can already induce bitflips within the refresh window in current DRAM chips.}}

\subsection{Disturbing DRAM Columns}
\label{sec:disturbing_columns}
{\figref{fig:bitflips_extended} \omcr{2}{in \secref{sec:intro}} shows the \X{}, RowHammer, RowPress, and retention failure bitflips from a single representative module \omcr{1}{(S0; Samsung 16Gb A-die)}.\ieycr{0}{\footnote{\label{fn:represent}We experiment with all \omcr{1}{\nMODULES{}} tested modules and observe \omcr{2}{very similar} characteristics \omcr{2}{to} this representative Samsung module\omcr{2}{, S0}.}} 
For RowHammer (RowPress), the same aggressor row (row 1536) is hammered (pressed, i.e., with \DRAMTIMING{AggOn}=70$\mu$s) for 16 seconds.} \ieycr{0}{For the retention failure test, the bank remains idle (i.e., in precharged state) for 16 seconds.} Dashed vertical lines indicate \omcr{1}{reverse-engineered} subarray boundaries: row addresses 0-1023 belong to Subarray 0, 1024-2047 to Subarray 1, and 2048-3071 to Subarray 2. The aggressor row {(row 1536)} is located in Subarray 1 (Aggressor Subarray), and Subarrays 0 and 2 are physically adjacent to the aggressor subarray (Neighboring Subarray).

\observation{\X{} induces bitflips across \omcr{2}{\emph{three}} consecutive subarrays (3072 rows), \omcr{1}{causing bitflips in many more rows than RowHammer and RowPress.}}

{\textbf{\em All \omcr{2}{3072} rows} in three consecutive subarrays experience ColumnDisturb bitflips. However, \emph{only} the immediate neighbor rows of the aggressor row 1536 (i.e., rows 1535 and 1537) experience RowHammer \& RowPress bitflips.}\footnote{{We \omcr{1}{verify} that bitflips \ieycr{0}{in the immediate neighbors of \ieycr{1}{an} aggressor row (+/-1 rows of the aggressor row)} are caused by RowHammer and RowPress \omcr{2}{based on three observations}. First, prior works on real DRAM chips~\cite{nam2023xray,nam2024dramscope,luo2023rowpress,luo2025revisiting} demonstrate that \ieycr{0}{by \omcr{2}{\emph{correctly}} reverse-engineering logical-to-physical row mapping of DRAM chips,} RowHammer \& RowPress induce bitflips in adjacent victim rows (+/-1 rows of the aggressor row)~\cite{luo2023rowpress,nam2023xray,nam2024dramscope,luo2025revisiting} due to \omcr{2}{the} underlying silicon-level characteristics \ieycr{0}{of RowHammer \& RowPress}~\cite{Zhou2024Unveiling, Zhou2024Understanding, luo2025revisiting}. Second, when we hammer different aggressor rows in the same subarray, we observe high bitflip counts \ieycr{1}{only} in  +1/-1 neighboring victim rows \ieycr{2}{(i.e., 7559 RowHammer bitflips and 5406 RowPress bitflips on average across +1 and -1 neighboring victim rows)}, while \ieycr{1}{all} other victim rows in the aggressor subarray \ieycr{0}{experience \ieycr{2}{similar} bitflip counts to each other \ieycr{2}{(between 2353-3505 \X{} bitflips across all other victim rows, significantly less than RowHammer \& RowPress bitflips}.} Third, when we initialize victim rows with an all-0 pattern, we \ieycr{1}{observe \ieycr{2}{only} \omcr{2}{0 to 1} bitflips} in \omcr{2}{the} +/-1 neighboring rows (not shown). \ieycr{1}{Since \X{} induces only 1 to 0 bitflips \omcr{2}{(\secref{sec:bitflip_direction})}, whereas  RowHammer~\&~RowPress can induce both 0 to 1 and 1 to 0 bitflips (\secref{sec:bitflip_direction}), \omcr{2}{this third observation} further indicates that RowHammer~\&~RowPress induce bitflips in only +/-1 rows.}}}

This observation could have serious implications for the future, as refresh-based DRAM read disturbance mitigation techniques consider \ieycr{0}{a few neighbor rows (e.g., up to +4/-4 in industry solutions~\cite{jedec2020ddr5,jedec2024jesd795c,hassan2021utrr,frigo2020trrespass,canpolat2024understanding,canpolat2025chronus})} of the aggressor row as victim rows\omcr{2}{, all in the same subarray}, \ieycr{0}{whereas \X{} leads to bitflips in \omcr{2}{\emph{thousands}} of rows (i.e., all rows across \omcr{2}{\emph{three}} consecutive subarrays)}.

We observe that subarrays that are neither the aggressor subarray nor physically-adjacent to the aggressor subarray \hluo{do \emph{not}} experience any \X{} bitflips (not shown in the figure). This is \omcr{0}{due to} the open-bitline \omcr{1}{DRAM architecture~\cite{dram-circuit-design,schloesser20086f,sekiguchi2002low,luo2020clrdram,chang2016lisa,jacob_book_2008,itoh2013vlsi,itoh77, lee2013tiered}}: \ieycr{0}{two neighboring subarrays share} half of \ieycr{0}{their columns. Thus, the aggressor subarray (Subarray 1) shares columns with both the above \omcr{1}{subarray} (Subarray 0) and \omcr{1}{the} below \omcr{1}{subarray} (Subarray 2) \ieycr{0}{(see \figref{fig:dram_organization})}}. As a result, \ieycr{0}{subarrays} that do not share \ieycr{0}{columns} with the aggressor subarray are not affected \ieycr{2}{(see \figref{fig:mech_intro}c)}. We hypothesize that this observation shows \X{} is a \ieycr{0}{column}-based read disturbance phenomenon, \omcr{0}{since} only the cells sharing the same \ieycr{0}{columns} as the aggressor row experience \X{} bitflips.  

\observation{\X{} induces more bitflips in \omcr{1}{the} aggressor subarray's rows than in those of neighboring subarrays.}
On average, \X{} induces \param{1.45}$\times$ more bitflips in the aggressor subarray\ieycr{1}{'s rows} than in the neighboring subarrays\ieycr{1}{' rows}. We hypothesize that \omcr{1}{all columns are affected in the aggressor subarray because an activation causes perturbation in all bitlines \omcr{2}{in the same subarray. In contrast, o}nly half of the columns in the neighboring subarray are connected to the aggressor subarray, and hence an activation causes a perturbation in \omcr{2}{\emph{half}} of the bitlines \omcr{2}{in the neighboring subarrays.}}

We also observe \ieycr{2}{no} overlap in the \hluo{\ieycr{0}{column} indices of the cells} that experience \emph{only} \X{} bitflips in Subarray 0 and Subarray 2 (neighboring subarrays of Subarray 1). {We hypothesize that this is because Subarray 1's even \ieycr{0}{columns} are shared with Subarray 0's odd \ieycr{0}{columns} (e.g., shared blue \ieycr{0}{columns} in the middle and top subarrays in \figref{fig:dram_organization}) and Subarray 1's odd \ieycr{0}{columns} are shared with Subarray 2's even \ieycr{0}{columns} (e.g., red \ieycr{0}{columns} in the middle and bottom subarrays in \figref{fig:dram_organization}).} This observation further supports our hypothesis that \X{} is a \ieycr{0}{column}-based read disturbance phenomenon \ieycr{0}{since \emph{only} cells that share columns with the aggressor row are affected}.

\observation{\label{obs:fig1_obs1} \omcr{1}{For a given refresh interval,} \X{} induces \omcr{1}{many} more bitflips than retention \ieycr{0}{failure\omcr{1}{s}}.}
On average, \X{} induces \param{2942.68} bitflips per row in the aggressor subarray and \param{2025.02} bitflips per row in the neighboring subarrays, which are \param{7.07}x and \param{4.87}x more than retention \ieycr{0}{failure \omcr{2}{bitflips} for a refresh interval of 16 seconds}, respectively.

\takeaway{\X{} is a column-based read disturbance phenomenon that disturbs cells \ieycr{0}{via DRAM columns} \omcr{1}{(i.e., bitlines)} and induces bitflips \omcr{2}{in} \ieycr{1}{thousands of rows (i.e.,} as many as \ieycr{1}{all \omcr{2}{3072 rows in}} three DRAM subarrays\ieycr{1}{)}.}

\subsection{Bitflip Direction}
\label{sec:bitflip_direction}

\begin{figure}[ht]
    \centering
    \includegraphics[width=\linewidth]{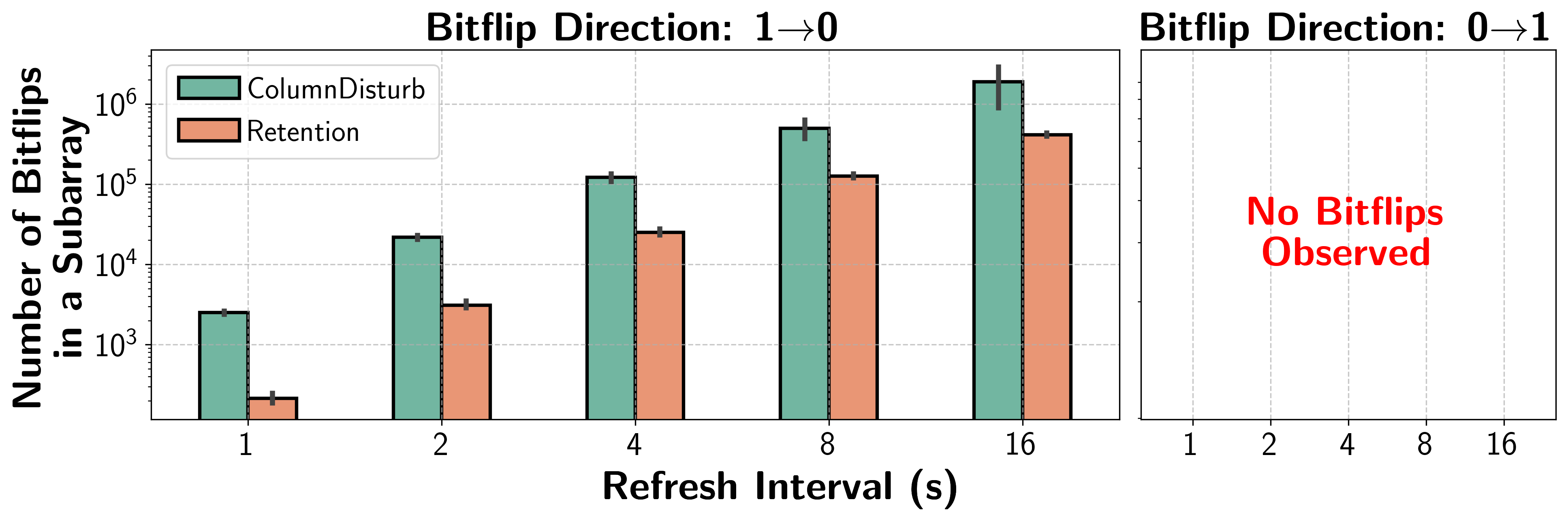}
    \caption{\hluo{\omcr{1}{D}istribution of the number of 1 to 0 and 0 to 1 bitflips in a subarray} \ieycr{0}{\omcr{1}{due to} \X{} and retention failures}.}
    \label{fig:bitflip_dir}
\end{figure}

We analyze the bitflip direction of \X{} \omcr{0}{versus} retention \ieycr{0}{failures} using one \ieycr{0}{representative} \omcr{2}{module} \omcr{1}{(S0; Samsung 16Gb A-die)}.\footref{fn:represent}
\figref{fig:bitflip_dir} shows the number of \hluo{1 to 0 (left) and 0 to 1 (right)} bitflips in a subarray (y-axis) 
\omcr{1}{due to \X{} and retention failure \omcr{2}{bitflips} across five different refresh intervals} (x-axis). Error bars represent the \hluo{range of minimum and maximum number of bitflips} across all tested subarrays. 

\observation{\label{obs:bit_dir1} \X{} induces \omcr{2}{\emph{only}} 1 to 0 bitflips, \omcr{0}{which are the} same \omcr{0}{direction} as retention \ieycr{0}{failures}.}
Across all tested \ieycr{2}{refresh intervals} and subarrays, the \omcr{2}{\emph{only}} \omcr{0}{observed} bitflip direction for \X{} is 1 to 0, which is the same as retention failures. \hluo{We do \emph{not} observe any 0 to 1 \omcr{2}{\X{}} bitflips\omcr{1}{. This is in stark contrast to} RowHammer and RowPress, \omcr{1}{which} induce bitflips in both \omcr{1}{1 to 0 and 0 to 1} directions~\cite{luo2023rowpress,nam2024dramscope,kim2020revisiting,kim2014flipping,luo2025revisiting}\ieycr{2}{, which we also observe in our tested DRAM chips}.}

\observation{\label{obs:bit_dir2} \X{} \omcr{1}{induces many more bitflips than} retention \ieycr{0}{failures}, especially at shorter \omcr{2}{refresh intervals}.}
On average, \X{} induces \param{11.77}x, \param{7.02}x, \param{4.86}x, \param{3.97}x, and \param{4.58}x more bitflips than retention \ieycr{0}{failures} at \omcr{1}{refresh intervals of} 1s, 2s, 4s, 8s, and 16s\omcr{1}{, respectively}.

\takeaway{DRAM is much more vulnerable to \X{} than retention \ieycr{0}{failures,} and the bitflip \hluo{directionalities} of \X{} and \ieycr{0}{RowHammer\omcr{1}{~\&~}RowPress are significantly different}.}

\subsection{\ieycr{0}{Data Pattern \omcr{2}{in Aggressor Row}}}
\label{sec:dp_column}

\begin{figure}[ht]
    \centering
    \includegraphics[width=\linewidth]{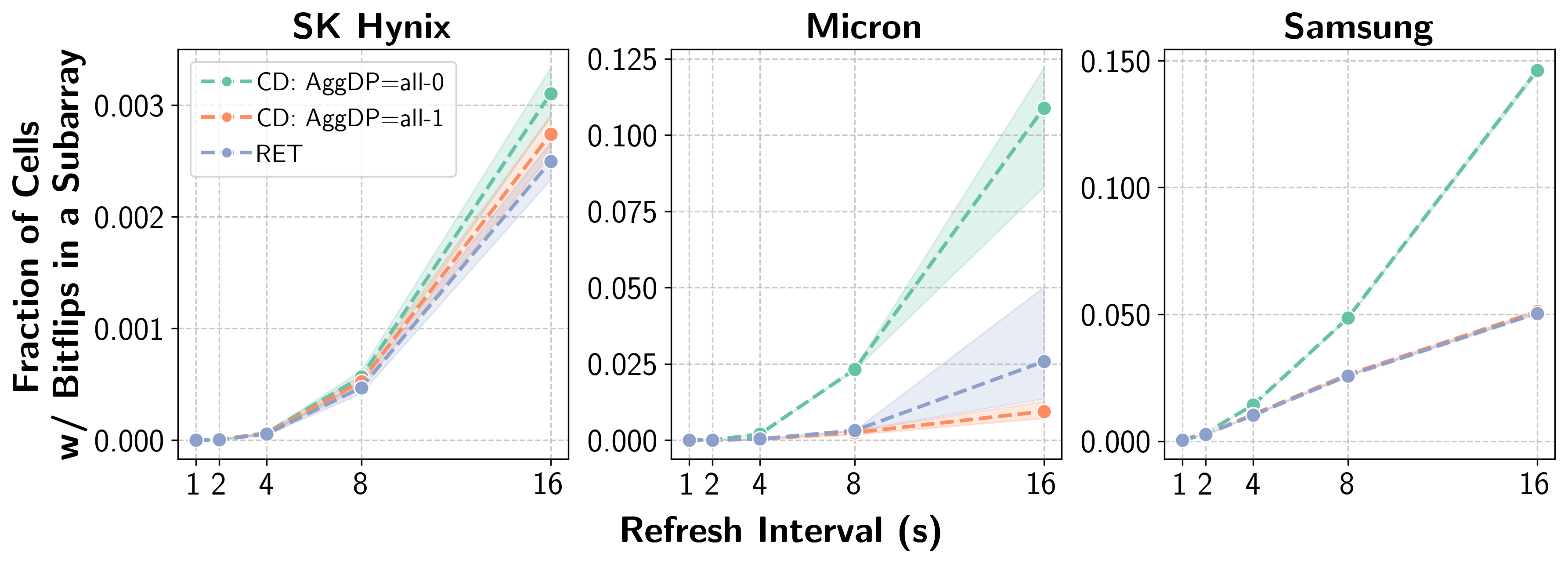}
    \caption{\ieycr{1}{Distribution of fraction of cells with bitflips in a subarray due to \X{} for two different aggressor data patterns and retention failures.} }
    \label{fig:agg_patt}
\end{figure}

We investigate how the \ieycr{0}{aggressor data pattern (i.e., data pattern on the columns \omcr{2}{perturbed by} the aggressor \omcr{2}{row})} affects the \X{} vulnerability using a single \ieycr{0}{representative} module from each \omcr{1}{major} DRAM manufacturer~\omcr{2}{(S0, H0, and M6)}. {W}\ieycr{0}{e test \omcr{1}{\emph{all}} subarrays in a single bank from each tested module}. For this experiment, we use all-1 as the data pattern \ieycr{0}{for all victim cells in a subarray \ieycr{1}{since \X{} induces \omcr{2}{only} 1 to 0 bitflips}} and \DRAMTIMING{AggOn}=$\tras{}$. \ieycr{0}{We observe that not all subarrays have the same number of rows (the number of rows in a subarray across all tested modules ranges between 512 and 1024), and thus, we use a metric \omcr{1}{\emph{"fraction of cells with bitflips in a subarray."}}}
\figref{fig:agg_patt} shows the fraction of cells that experience bitflips in a subarray \ieycr{1}{across all tested subarrays} (y-axis) \omcr{2}{for} five different \ieycr{1}{refresh intervals} (x-axis) for three DRAM manufacturers (subplot). Each line represents a different \ieycr{0}{experiment}: \ieycr{0}{1) \X{}: AggDP=all-0, \omcr{2}{the} aggressor row is initialized with all-0, 2) \X{}: AggDP=all-1, \omcr{2}{the} aggressor row is initialized with all-1, and 3) RET, retention failures, where all columns are at the precharged voltage (VDD/2) during the \omcr{2}{entire} experiment. \ieycr{1}{The error band of a line shows the minimum and maximum values across all tested subarrays.}}

\observation{\label{obs:agg_patt1} \X{} induc\ieycr{1}{es many more} bitflips when the \ieycr{0}{aggressor \omcr{2}{row} data pattern is all-0}~\omcr{3}{versus} \ieycr{0}{all-1}.}
For example, at 16s, \ieycr{0}{all-0 aggressor pattern} induces \param{1.15}x, \param{11.52}x, and \param{2.86}x more bitflips than the \ieycr{0}{ all-1 aggressor data pattern} for \X{} in SK Hynix, Micron, and Samsung \omcr{2}{chips}, respectively.
We hypothesize that this occurs due to a larger voltage level difference between the victim cell and the \ieycr{0}{\ieycr{2}{perturbed columns (i.e., columns connected to the aggressor row)}. This is because, when the aggressor data pattern is all-0, all \ieycr{2}{perturbed} columns in the subarray are at GND (i.e., low voltage) while the victim cells are at VDD, resulting in a VDD difference. In contrast, when the aggressor data pattern is all-1, both victim cells and all \ieycr{2}{perturbed} columns in a subarray are at VDD, resulting in no voltage difference. We} further analyze the relationship between the \ieycr{0}{\ieycr{2}{perturbed} column} voltage level and \X{} in \secref{sec:avg_bl_voltage}.

\observation{\label{obs:agg_patt2} \ieycr{1}{When the aggressor data pattern is all-1 (same as victim data pattern), \X{} can induce fewer bitflips than retention failures.}}

For example, in Micron, \ieycr{1}{with a refresh interval of} 16s, \param{2.73}x fewer DRAM cells in a subarray on average experience bitflips with \X{} (\ieycr{1}{when both the aggressor data pattern and the victim data pattern are all-1}) \omcr{3}{versus} retention \ieycr{0}{failure}s.

We hypothesize that \ieycr{2}{this could be} due to the reduced voltage level difference between a \ieycr{0}{column} and \hluo{the victim} cell. 
\hluo{In \X{} \ieycr{0}{with all-1 aggressor \ieycr{1}{and victim} data pattern tests}, the voltage difference between the \ieycr{0}{victim} cell (VDD) and the \ieycr{0}{column} (VDD) is zero.} In contrast, \hluo{in retention failure tests, the \ieycr{0}{column}} is held at VDD/2 while the cell remains at VDD, resulting in a higher \ieycr{1}{column-cell} voltage difference (i.e., VDD/2) \ieycr{1}{than \X{} with all-1 aggressor and victim data pattern tests. As a result, having a lower voltage difference between the victim cell and the column could lead to a lower \ieycr{2}{\X{} vulnerability.}}
We provide a detailed analysis and hypotheses for the effect of \ieycr{2}{the} voltage difference \ieycr{2}{between victim cells and perturbed columns} in \secref{sec:avg_bl_voltage}.

\takeaway{\ieycr{0}{Data pattern \ieycr{2}{in the aggressor row}} has a significant effect on the \omcr{3}{number of observed} \X{} bitflips.}

\subsection{Aggressor Row on Time}
\label{sec:tAggON}
\begin{figure}[ht]
    \centering
    \includegraphics[width=\linewidth]{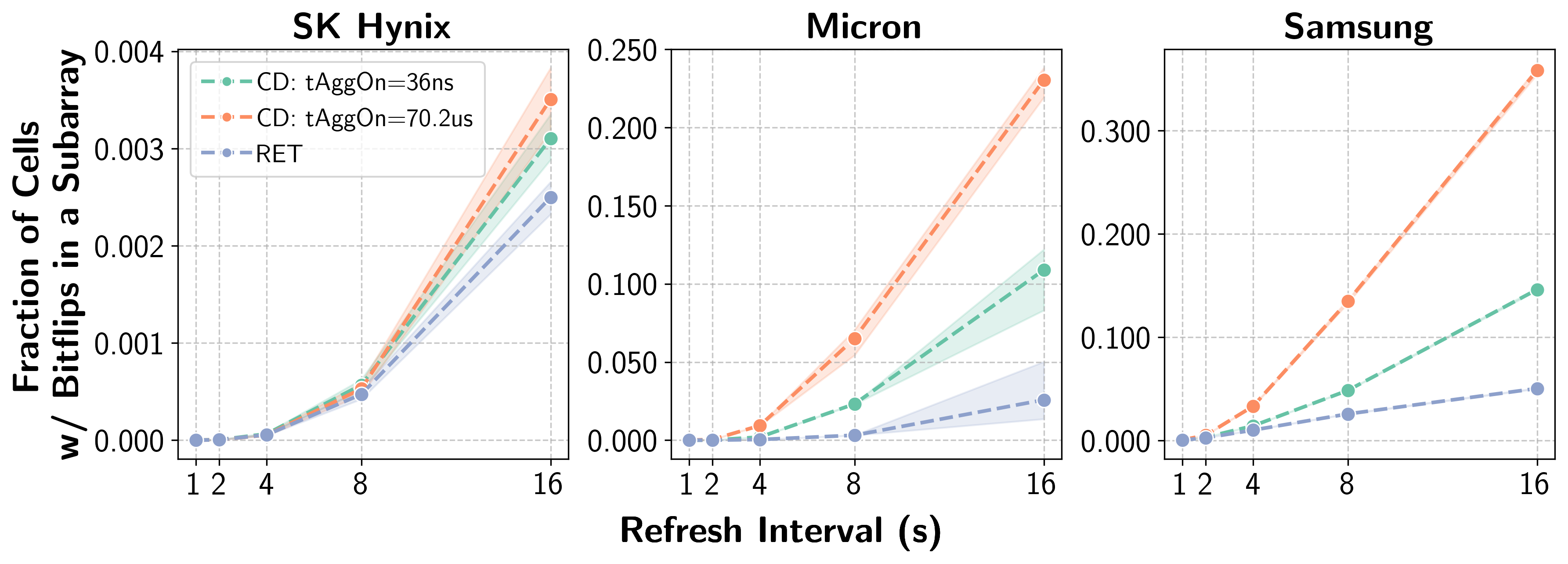}
    \caption{\ieycr{1}{Distribution of fraction of cells with bitflips in a subarray due to \X{} for two different \DRAMTIMING{AggOn} values and retention failures.}}
    \label{fig:taggon}
\end{figure}

We study the effect of the amount of time an aggressor row is kept open (\DRAMTIMING{AggOn}) on \X{} \omcr{2}{bitflips}.\footnote{\ieycr{1}{Keeping an aggressor row open for a long time (i.e., high \DRAMTIMING{AggOn} values) perturbs \omcr{2}{\emph{all}} columns connected to the cells in the aggressor row, since the columns remain driven to VDD or GND.}}
For this analysis, we use all-1 (0xFF) as the victim data pattern and all-0 (0x00) for the aggressor data pattern \ieycr{0}{since these are the worst-case data patterns that induce the most \X{} bitflips in a DRAM chip.} \figref{fig:taggon} shows the fraction of cells with bitflips in a subarray \omcr{1}{observed in} three \omcr{1}{representative} modules, one from each tested manufacturer. Each line represents a different \ieycr{0}{experiment: 1) \X{}: \DRAMTIMING{AggOn}=36ns, where we keep the aggressor row open for 36ns, 2) \X{}: \DRAMTIMING{AggOn}=70.2$\mu$s, where we keep the aggressor row open for 70.2$\mu$s, and 3) RET, retention failure, where the bank is idle (i.e., \omcr{2}{in} precharged state) during the experiment.} 

\observation{\label{obs:taggon_1} \ieycr{0}{Fraction of cells with \X{} bitflips in a subarray significantly increases as} \DRAMTIMING{AggOn} increases.}
For example, \ieycr{0}{at a refresh interval of} 16s, increasing \DRAMTIMING{AggOn} from 36ns to 70.2$\mu$s increases the number of \X{} bitflips by \param{1.20}x, \param{2.12}x, and \param{2.45}x for SK Hynix, Micron, and Samsung modules, respectively. We hypothesize that this occurs because a longer \DRAMTIMING{AggOn} keeps the \ieycr{3}{perturbed} column\omcr{1}{s} at a low voltage level (GND) for an extended period, which \omcr{1}{likely} result\omcr{1}{s} in an increased \X{} effect.

\subsection{Average Voltage Level on \ieycr{0}{\omcr{2}{Perturbed} Column\omcr{1}{s}}}
\label{sec:avg_bl_voltage}
\ieycr{0}{We observe that DRAM chips become significantly more vulnerable to \X{} 1) when the voltage level difference between the victim cell and the \omcr{2}{perturbed} columns (i.e., columns \omcr{2}{perturbed by} the aggressor \omcr{2}{row}) is high (\secref{sec:dp_column}) and 2) when the \omcr{2}{perturbed} columns are kept active longer (\secref{sec:tAggON}), i.e., the voltage level difference between the victim cell and the \omcr{2}{perturbed} column\omcr{1}{s is kept} high for an extended period.} Based on our observations in \secref{sec:dp_column} and \secref{sec:tAggON}, we hypothesize that the voltage level on \omcr{1}{the} \ieycr{0}{\omcr{2}{perturbed} column}\omcr{1}{s} is a key parameter that affects the \X{} vulnerability in DRAM. 

To understand the effect of average column voltage level \omcr{1}{on \X{}}, we \ieycr{0}{calculate the average voltage level on the \omcr{2}{perturbed} column\omcr{1}{s} $AVG(V_{COL})$ when we perform \X{} \ieycr{1}{tests}. Our access pattern, \omcr{1}{for} every \DRAMTIMING{AggOn}+$\trp{}$ time, (\secref{subsec:method}) 1) keeps the column voltage level at the aggressor data pattern in the aggressor subarray, $DP_{COL}$, for \DRAMTIMING{AggOn} and 2) keeps the column voltage level at precharged voltage (VDD/2) for $\trp{}$.}

\ieycr{0}{Based on the relationship between $AVG(V_{COL})$ and $DP_{COL}$, \DRAMTIMING{AggOn}, VDD/2, and $\trp{}$ parameters, we can calculate $AVG(V_{COL})$ as follows:}

$AVG(V_{COL}) = \frac{\DRAMTIMING{AggOn}*DP_{COL} + VDD/2*\trp{}}{\DRAMTIMING{AggOn} + \trp{}}$

\ieycr{0}{For example, assume $DP_{COL}$=GND=0, \DRAMTIMING{AggOn}=36ns, precharged voltage level=VDD/2, and $\trp{}$=14ns (i.e., \omcr{2}{perturbed} columns are at GND for 36ns, and at VDD/2 for 14ns in every 50ns), $AVG(V_{COL})$ becomes $\frac{36ns*0 + VDD/2*14ns}{36ns + 14ns}=0.14*VDD$.}

\begin{figure}[ht]
    \centering
    \includegraphics[width=\linewidth]{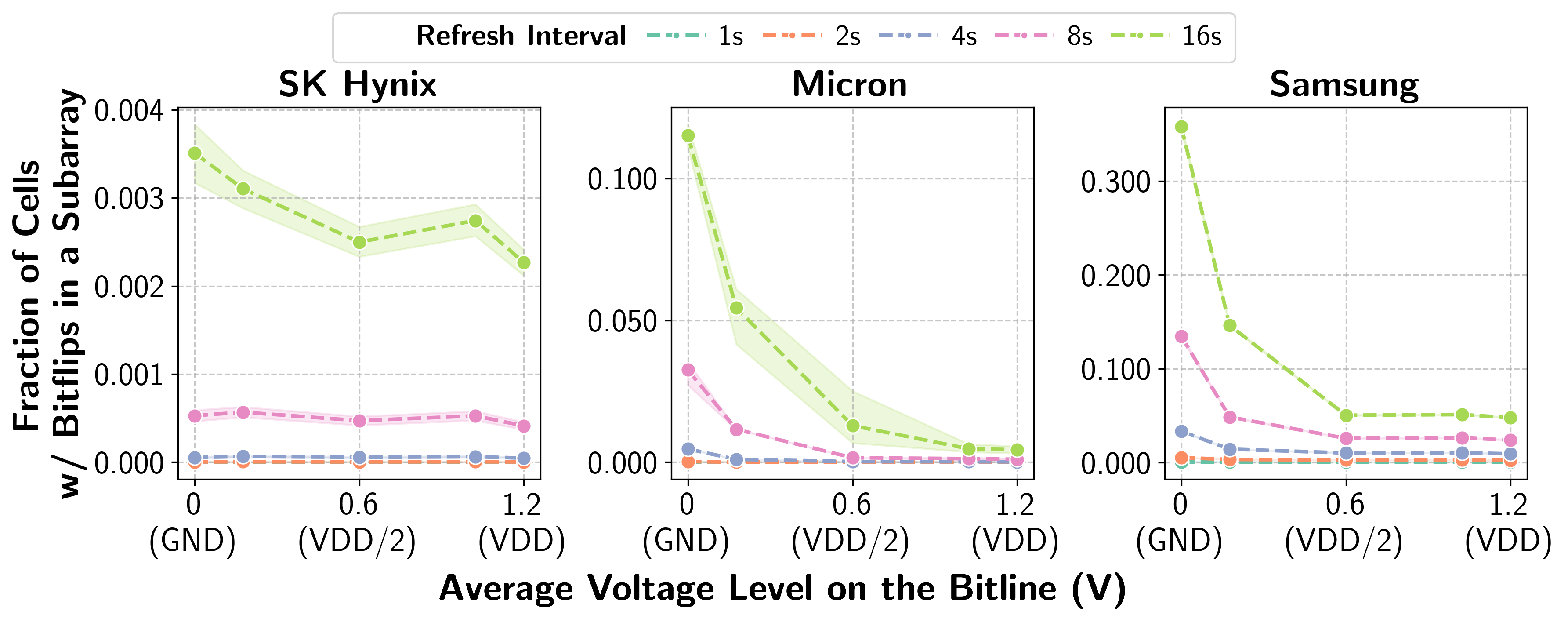}
    \caption{\hluo{\omcr{0}{F}raction of cells with \X{} bitflips for different average \ieycr{0}{column} voltage levels}.}
    \label{fig:column_voltage}
\end{figure}

\figref{fig:column_voltage} shows the fraction of cells with bitflips (y-axis) as we sweep the average voltage level on the \ieycr{2}{perturbed} \ieycr{0}{columns} (x-axis) \ieycr{1}{across} five refresh interval\omcr{1}{s} (each colored line). \ieycr{0}{In this experiment, we test all subarrays in a single bank from three DRAM modules, one from each tested manufacturer.}

\observation{\label{obs:voltage_1} \ieycr{0}{A DRAM chip} becomes significantly \omcr{1}{\emph{more vulnerable}} to \X{} as the average voltage level o\ieycr{0}{n} the \ieycr{0}{\omcr{2}{perturbed} column}\ieycr{1}{s} \omcr{1}{\emph{decreases}}.}
We observe that reducing the average \ieycr{0}{column} voltage level from VDD to GND increases the fraction of cells that experience bitflips in a subarray by \param{1.65}x, \param{26.31}x, and \param{7.50}x for SK Hynix, Micron, and Samsung, respectively, at \ieycr{0}{a refresh interval of} 16 seconds. 

\hluo{We provide two hypotheses to explain our observations. As the \ieycr{0}{\omcr{2}{perturbed} column} voltage decreases, the voltage difference across the victim cell and the \ieycr{0}{\omcr{2}{perturbed} column} increases, which 1) exacerbates subthreshold leakage of the access transistor~\cite{Roy2003Leakage} and/or 2) exacerbates the dielectric leakage between the \omcr{0}{victim} capacitor and the \ieycr{0}{\omcr{2}{perturbed} column}~\cite{Yu2022TheStudy}.}\footnote{We call for future device-\omcr{0}{level} and silicon-level studies to develop a better understanding of the inner workings of \X{}\omcr{0}{, just} as device-level studies (e.g., \cite{Zhou2024Unveiling, Zhou2024Understanding}) did for RowPress \emph{after} the RowPress paper~\cite{luo2023rowpress} demonstrated for the RowPress phenomenon \omcr{0}{experimentally on real DRAM chips}.}
\hypothesis{{\X{} exacerbates subthreshold leakage of the access transistor~\cite{Roy2003Leakage} and/or the dielectric leakage between the capacitor and the bitline~\cite{Yu2022TheStudy}.}}

\takeaway{The voltage level on the \ieycr{0}{column} plays an important role in \X{}'s device-level failure mechanisms\omcr{0}{. D}evice-level investigation is necessary to develop a better \omcr{0}{fundamental \omcr{1}{and} first-principles} understanding of \X{}.}

\subsection{Blast Radius}
\label{sec:blast_radius}
\ieycr{1}{We study how many rows experience \X{} bitflips in a subarray for a given refresh interval. To do so, we use a metric\omcr{2}{:} \emph{the number of rows with bitflips in a subarray}, i.e., the number of rows that experience at least one bitflip in a subarray, broadly referred to as \omcr{2}{\emph{blast radius}}~\cite{kim2014flipping,kim2020revisiting,orosa2021deeper,yaglikci2022understanding,luo2023rowpress, hassan2021utrr, yaglikci2022hira, yaglikci2024svard, olgun2025variable, tugrul2025understanding,lang2023blaster,kogler2022half}.}

\begin{figure}[ht]
    \centering
    \includegraphics[width=\linewidth]{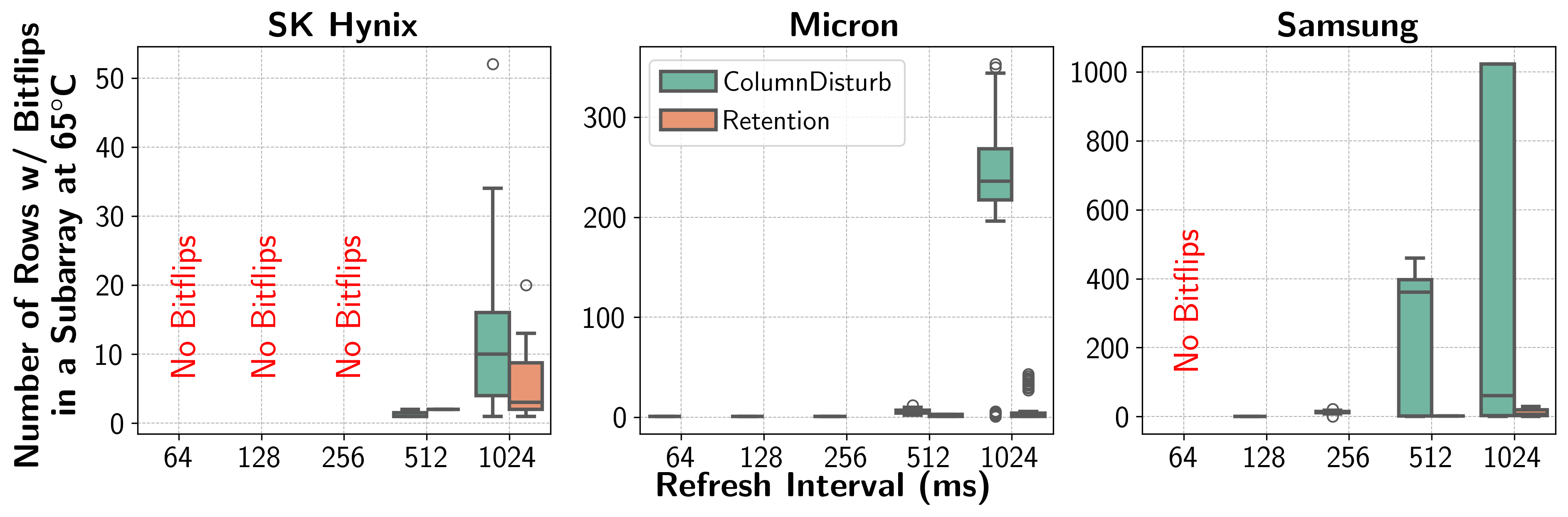}
    \caption{\hluo{Distribution of the number of rows that experience \X{} \omcr{0}{versus} retention \ieycr{0}{failure} bitflips for \ieycr{0}{different \omcr{2}{refresh intervals}}}.}
    \label{fig:blast}
\end{figure}

\figref{fig:blast} shows the distribution of the number of rows \ieycr{1}{with bitflips} in a subarray at 65$^{\circ}$C as a boxplot. Each subplot is dedicated to a different manufacturer and shows how many rows experience \X{} and retention failure bitflips in a subarray (y-axis) for five different \ieycr{2}{refresh intervals}.

\observation{\label{obs:blast_1} \X{} induces bitflips \omcr{0}{in} significantly more DRAM rows than retention \ieycr{1}{failures}.}
For example, with a refresh interval of 1024ms, \X{} (Retention) induces bitflips in up to \param{52} (\param{20}), \param{353} (\param{34}), and \param{1022} (\param{29}) rows for SK Hynix, Micron, and Samsung modules, respectively. For \ieycr{0}{SK Hynix, Micron, and} Samsung modules, \ieycr{0}{respectively, with a refresh interval of 512ms}, \X{} induces bitflips in \ieycr{0}{2, 6, and} \param{232} rows \emph{on average}, whereas only two rows experience retention failures \emph{at most}. 

\observation{\label{obs:blast_2} \omcr{1}{Number of rows with} \X{} \omcr{1}{bitflips} significantly increases as the \ieycr{1}{refresh interval} increases.}
For example, from 512ms to 1024ms, for SK Hynix, Micron, and Samsung, respectively, \param{9.73}, \param{215.85}, and \param{159.22} more rows \ieycr{1}{in a subarray} experience \X{} bitflips\omcr{0}{,} on average \ieycr{1}{across all tested subarrays}. However, for retention \omcr{2}{failures}, the increase is significantly smaller than \X{}. From 512ms to 1024ms, up to \param{8.39} \ieycr{1}{(\param{2.76} on average)} more rows experience retention \ieycr{0}{failure} bitflips on average \ieycr{0}{across all tested modules}.

\takeaway{Significantly more rows are vulnerable to \X{} than retention \ieycr{0}{failures.}}

\subsection{\X{} on COTS HBM Chips}
\label{sec:hbm}
\begin{figure}[ht]
    \centering
    \includegraphics[width=0.94\linewidth]{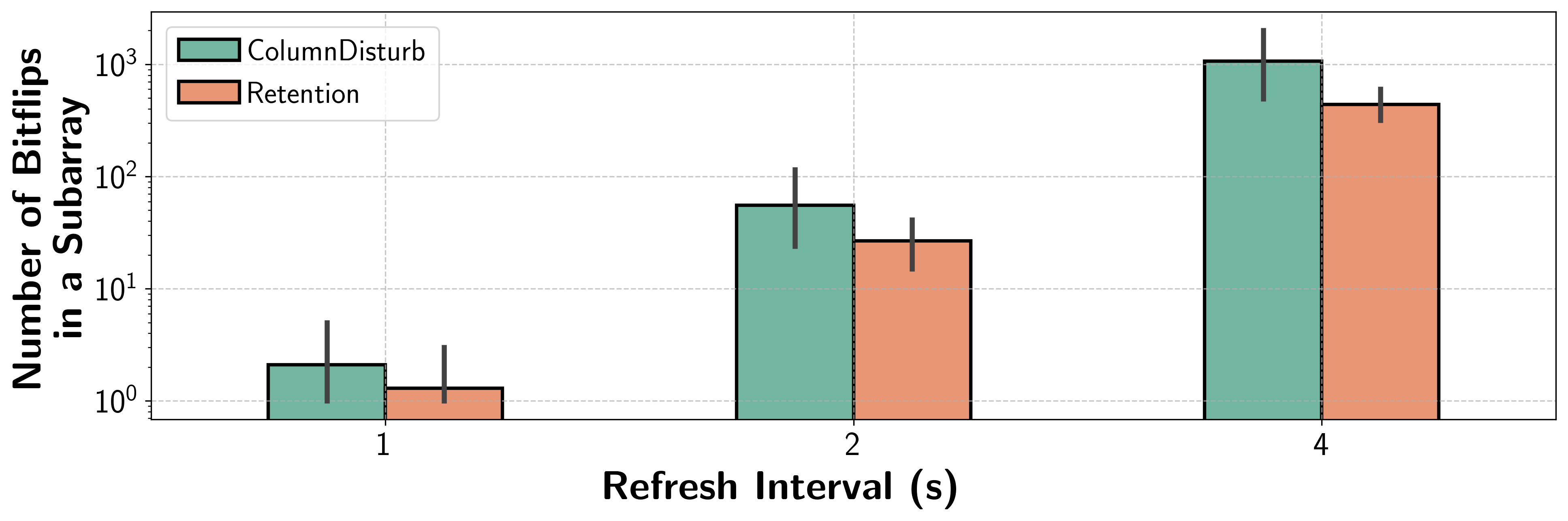}
    \caption{\hluo{Number of \X{} and retention failure bitflips for} \ieycr{0}{different} \ieycr{1}{refresh intervals in HBM2 chips}.}
    \label{fig:hbm_bitflip}
\end{figure}

So far, we characterize commodity DDR4 chips to \omcr{1}{demonstrate and} understand \X{}. In this section, we provide a preliminary study on the vulnerability of real HBM2 chips to \X{}. We test 4 HBM2 chips and analyze the number of \X{} bitflips and retention failures across all subarrays in a bank from three HBM2 chips.
\figref{fig:hbm_bitflip} shows the number of bitflips in a subarray across \ieycr{0}{all subarrays from a single bank} when 1) performing \X{} and 2) keeping the bank in the idle state without performing any refresh (i.e., retention)\omcr{0}{,} for 1, 2, and 4 seconds. \ieycr{1}{Error bars represent the \hluo{range of minimum and maximum number of bitflips} across all tested subarrays.}

\observation{HBM2 chips are vulnerable to \X{}, and \X{} induces many more bitflips than retention failures.}
We observe that across all \omcr{0}{four} HBM2 chips tested, \X{} induces \param{1.61}x, \param{2.08}x, and \param{2.43}x more bitflips than retention failures in 1s, 2s, and 4s, respectively. \omcr{2}{We expect our \omcr{3}{other} observations for DDR4 chips (\secref{sec:foundational} and \secref{sec:indepth}) will hold for HBM2 chips as well, since the DRAM array is the same for both, and \X{} happens at the array level.}

\takeaway{\X{} is a widespread read disturbance phenomenon \omcr{1}{in DRAM chips}: not only DDR4 chips, but also HBM2 chips are vulnerable to \X{}.}

\section{In-Depth \X{} Analysis}
\label{sec:indepth}
\omcr{0}{T}his section further enhance\omcr{0}{s} our analysis of \X{} by investigating
parameters that have been shown to impact \omcr{0}{both} read disturbance \omcr{0}{and retention failure} mechanisms: temperature (\secref{sec:id_temp}), aggressor row on time (\secref{sec:id_taggon}), access pattern (\secref{sec:id_acc_patt}), data pattern (\secref{sec:id_dp}), the location of the aggressor row in a subarray (\secref{sec:id_loc}), and the effectiveness of Error Correcting Codes (ECC) against \X{} (\secref{sec:id_ecc}).\footnote{We observe that the overall trends are consistent across the three evaluated metrics: 1) time to induce the first \X{} bitflip in a subarray, 2) the fraction of cells with \X{} bitflips in a subarray, and 3) the blast radius. Due to space limitations, we provide the analysis of the first metric in all experiments and provide a detailed analysis of all metrics only for temperature. \omcr{0}{Our extended version provides comprehensive analyses of all metrics~\cite{yuksel2025columndisturb}}.} We conduct our experiments by testing all subarrays in all banks from all tested modules using the parameters \omcr{0}{with which} a DRAM chip experiences the highest \X{} vulnerability (i.e., aggressor data pattern=all-0, victim data pattern=all-1, \DRAMTIMING{AggOn}=70.2$\mu$s, temperature= 85$^{\circ}$C)\omcr{0}{,} unless stated otherwise.

\subsection{Temperature}
\label{sec:id_temp}
We analyze the relation\omcr{0}{ship} between temperature and the \X{} vulnerability by evaluating three metrics for four different temperature levels\omcr{0}{: 45$^{\circ}$C, 65$^{\circ}$C, 85$^{\circ}$C, and 95$^{\circ}$C}.

\begin{figure}[ht]
    \centering
    \includegraphics[width=\linewidth]{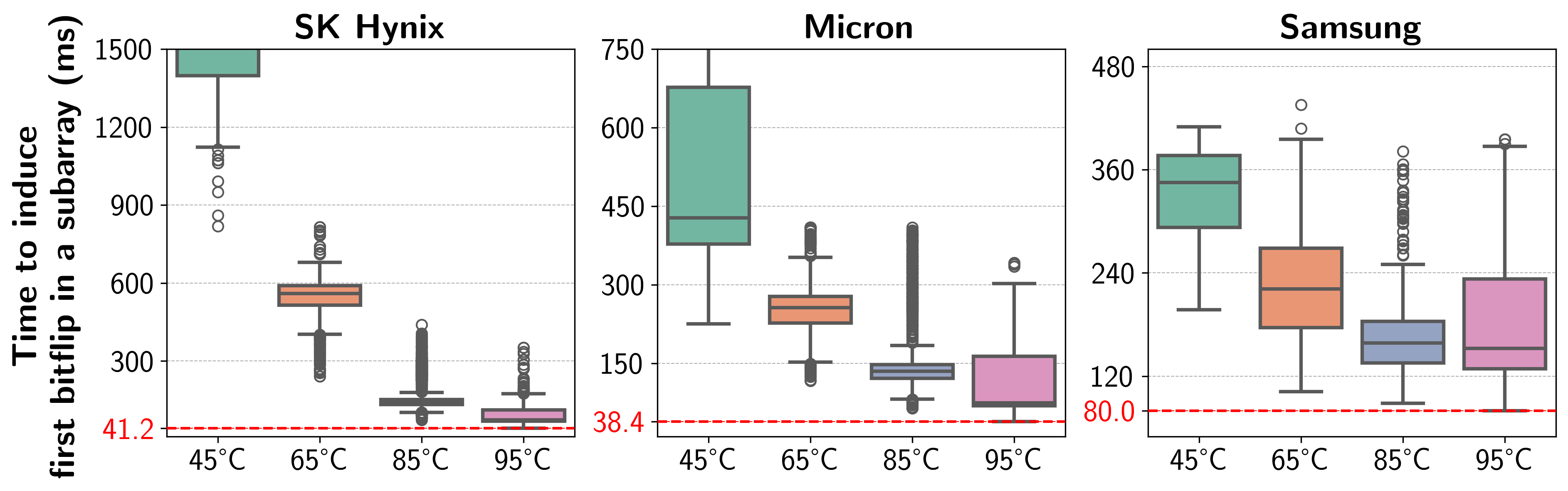}
    \caption{Distribution of time to induce the first \X{} bitflip at different temperatures.}
    \label{fig:hcfirst_temp}
\end{figure}

\noindent\textbf{Time to Induce the First \omcr{0}{\X{}} Bitflip.} \figref{fig:hcfirst_temp} shows how the time to induce the first bitflip changes with temperature using a box and whiskers plot. Each subplot is dedicated to a different manufacturer.

\observation{Time to induce the first \X{} bitflip significantly reduces as temperature increases.}
Across all tested modules, time to induce the first \X{} bitflip consistently reduces as temperature increases. For example, increasing temperature from 45$^{\circ}$C to 95$^{\circ}$C reduces the average time to induce the first \X{} bitflip in a subarray by \param{9.05}x, \param{5.15}x, and \param{1.96}x for SK Hynix, Micron, and Samsung, respectively.

\begin{figure}[ht]
    \centering
    \includegraphics[width=\linewidth]{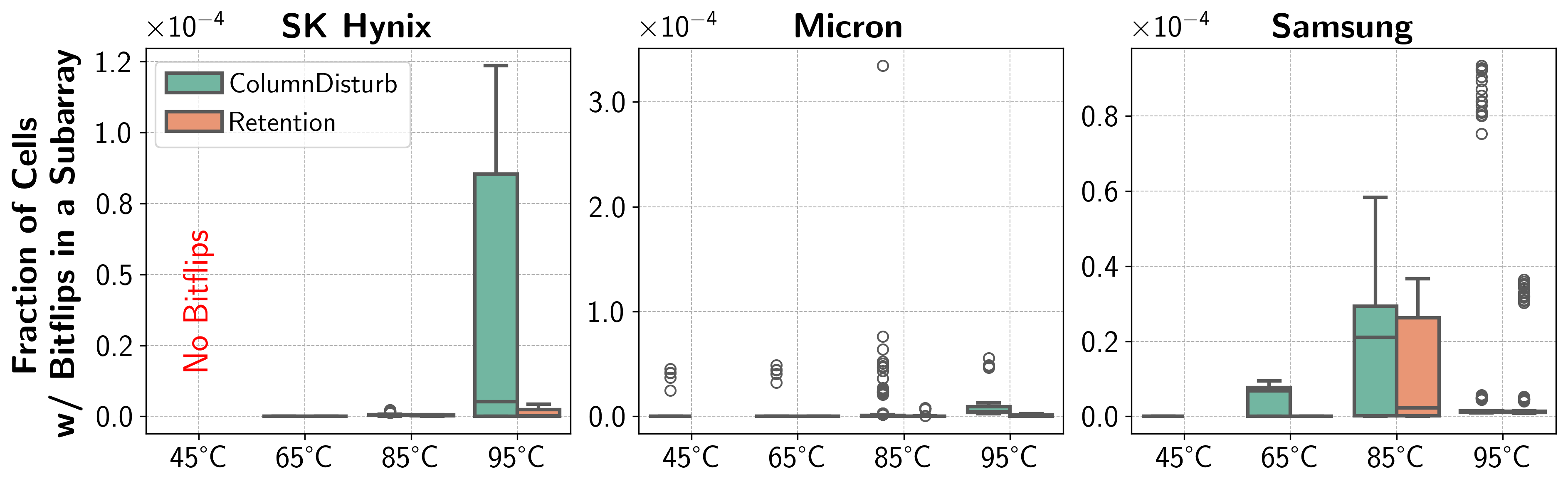}
    \caption{Distribution of fraction of cells with \X{} and retention failure bitflips \ieycr{0}{in a subarray} at different temperatures.}
    \label{fig:ber_temp}
\end{figure}

\noindent\textbf{Fraction of Cells with Bitflips in a Subarray.} \figref{fig:ber_temp} shows the fraction of cells with bitflips in a subarray (y-axis) \omcr{0}{due to} \X{} \omcr{0}{versus} retention failure\omcr{0}{s} (color-coded) across four temperature levels (x-axis) when \omcr{0}{the refresh interval} is 512ms.

\observation{\label{obs:ber_temp} \X{} \omcr{0}{is} more sensitive to temperature than retention failures.}
For example, in SK Hynix chips, when the temperature increases from 85$^{\circ}$C to 95$^{\circ}$C, \X{} \ieycr{0}{(retention failure) bitflips increase by} \param{72.96}x (\param{3.68}x) on average. We also observe that \X{} induces more bitflips than retention failure\omcr{0}{s} in all tested temperature levels across all DRAM modules. For example, at 65$^{\circ}$C, \X{} induces \param{152.66}x more bitflips than retention failure\omcr{0}{s} in Samsung modules.

\begin{figure}[ht]
    \centering
    \includegraphics[width=\linewidth]{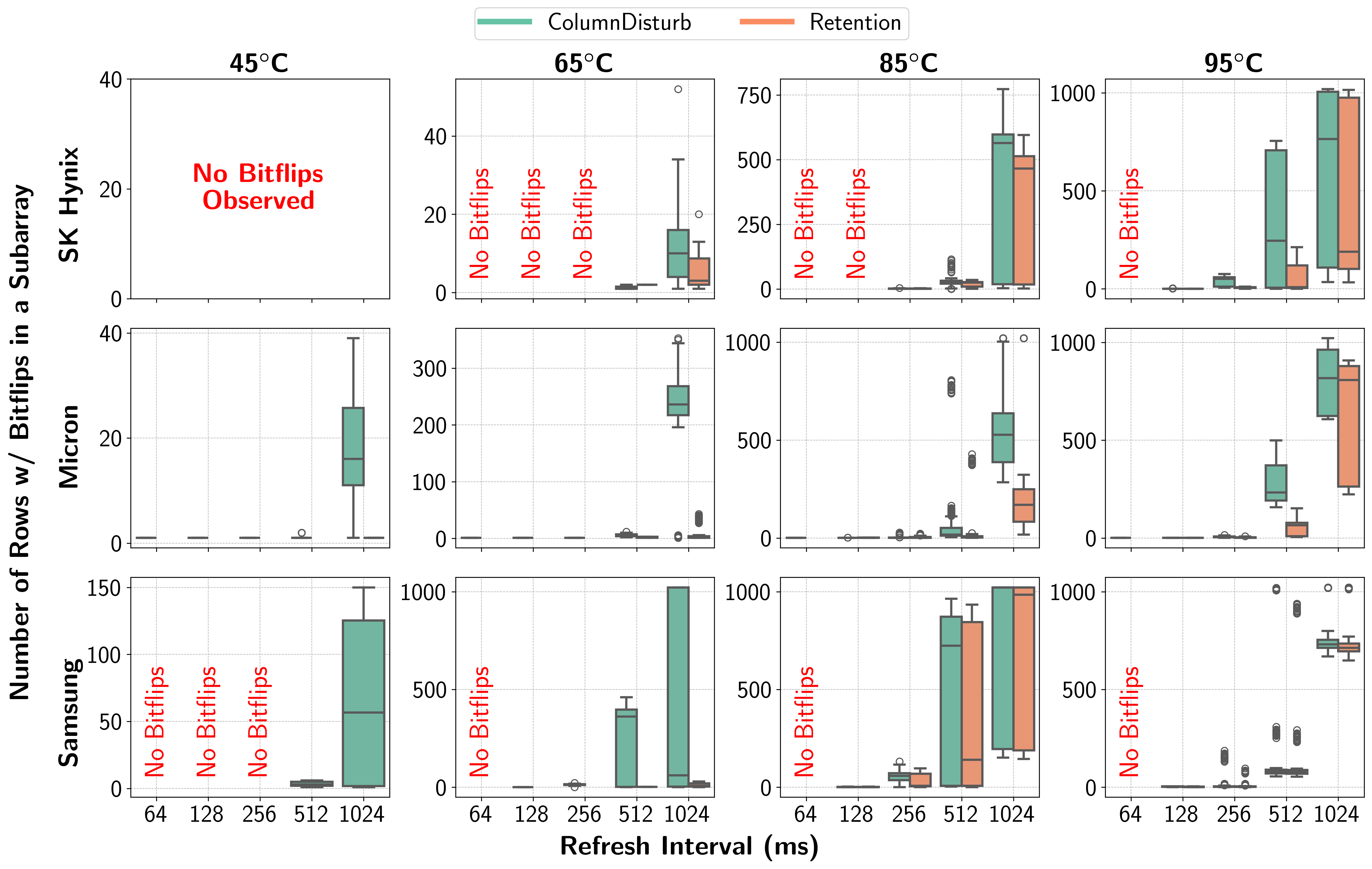}
    \caption{\ieycr{1}{Number of rows that experience at least one bitflip (i.e., blast radius) due to} \X{} and retention \omcr{0}{failures} for \ieycr{0}{different refresh intervals and temperature levels}.}
    \label{fig:blast_temp}
\end{figure}

\noindent\textbf{Blast Radius.} \figref{fig:blast_temp} shows the blast radius effect (y-axis) of \X{} and retention \omcr{0}{failures} \ieycr{0}{(i.e., the number of rows in a subarray that experience at least one \X{} and retention failure bitflip)} for three manufacturers (rows of subplots) and four temperature levels (columns of subplots). The x-axis shows the refresh interval in milliseconds (ms).

\observation{\label{obs:blast_temp1} \X{} induce\omcr{0}{s} bitflips in many more rows than retention failures \ieycr{0}{in all tested temperature levels}.}
For example, \omcr{1}{even} at \omcr{1}{a low temperature level,} \param{45}$^\circ$C, in Micron and Samsung, respectively, \X{} induces bitflips in up to \param{39} and \param{150} rows \ieycr{0}{in a subarray across all tested subarrays}. However, at most only a single row exhibits retention failures for Micron, and none for Samsung. We observe that across all tested temperature levels and refresh intervals, \X{} induces bitflips in up to 198x more DRAM rows
than retention failures\ieycr{0}{.}

\observation{\label{obs:blast_temp2} \omcr{0}{T}he blast radius \omcr{0}{of} both \X{} and retention \omcr{0}{failures increase with higher temperature}.}
At 95$^{\circ}$C, \ieycr{1}{across all tested subarrays,} both mechanisms nearly span an entire subarray (i.e., \ieycr{0}{almost} all rows \ieycr{0}{of every tested} subarray experience bitflips), whereas \X{} begins exhibiting a wide impact already at 65$^{\circ}$C (e.g., at least 5.33\% \ieycr{1}{of} subarrays in Samsung modules experience \omcr{0}{a} blast radius of \param{1000}).

\takeaway{\omcr{0}{As temperature increases, DRAM chips become more vulnerable to \X{}.}}

\subsection{Aggressor Row On Time}
\label{sec:id_taggon}
\begin{figure}[ht]
    \centering
    \includegraphics[width=0.95\linewidth]{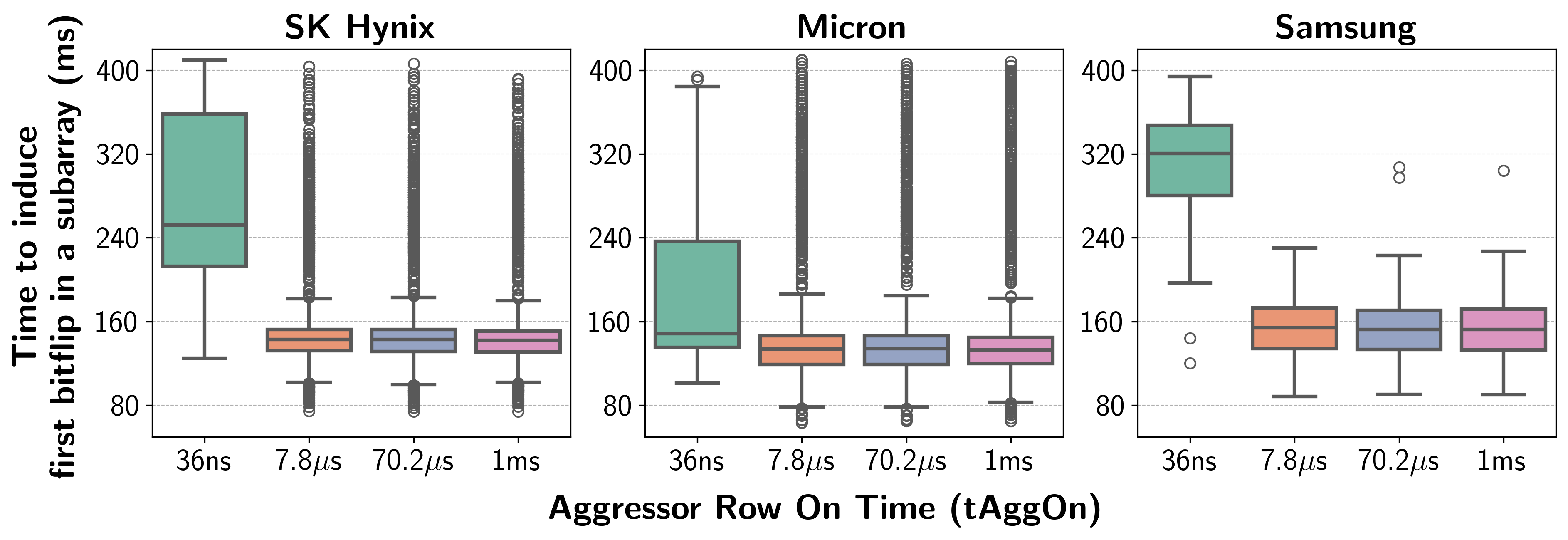}
    \caption{Time to induce first \X{} bitflip distribution \omcr{0}{for four different} \DRAMTIMING{AggOn} values.}
    \label{fig:hcfirst_taggon}
\end{figure}

\figref{fig:hcfirst_taggon} shows the distribution of time to induce the first \X{} bitflip in a subarray across all tested subarrays for four different \DRAMTIMING{AggOn} values: 36ns, 7.8$\mu$s, 70.2$\mu$s, and 1ms.

\observation{\omcr{0}{Keeping a row open} (\DRAMTIMING{AggOn} >\ieycr{0}{>} $\tras{}$) is significantly more effective than \omcr{0}{immediately closing it} (\DRAMTIMING{AggOn} = $\tras{}$) in inducing the first \X{} bitflip in a subarray.}
Increasing \DRAMTIMING{AggOn} from 36ns to 7.8$\mu$s reduces the average time to induce the first \X{} bitflip by \param{1.68}x, \param{1.22}x, and \param{2.03}x in SK Hynix, Micron, and Samsung, respectively. We observe that when \DRAMTIMING{AggOn} >\ieycr{0}{>} $\tras{}$, the distributions are very similar across all tested \DRAMTIMING{AggOn} \ieycr{1}{values}. We hypothesize that \ieycr{0}{increasing \DRAMTIMING{AggOn} \omcr{1}{decreases} the average voltage level on DRAM columns}, \ieycr{0}{resulting in higher \X{} vulnerability in DRAM chips (i.e., less time to induce the first \X{} bitflip in a subarray),} as discussed in \secref{sec:avg_bl_voltage}.

\subsection{Access Pattern}
\label{sec:id_acc_patt}
Until now, we perform \X{} with a single aggressor row, causing the column voltage to alternate between VDD/2 and GND or VDD, depending on the aggressor data pattern (described in \secref{subsec:method}). To understand the effect of toggling the column, we introduce a two-aggressor access pattern where two aggressor \omcr{0}{rows are used with} complementary data patterns: 

\act{} $R_{Agg1}$ $\xrightarrow[]{t_{Agg_{On}}}$ \pre{} $\xrightarrow[]{t_{RP}}$ \act{} $R_{Agg2}$ $\xrightarrow[]{t_{Agg_{On}}}$ $\cdots$ 

In the single aggressor access pattern (\secref{subsec:method}), we use all-0 as the aggressor data pattern. Thus, the column voltage transition is
\{GND$\rightarrow$VDD/2$\rightarrow$GND$\cdots$\}. 
\omcr{0}{In} the two\omcr{0}{-}aggressor \omcr{0}{access pattern}, we \omcr{0}{use} all-0 in the first aggressor ($R_{Agg1}$) and all-1 in the second aggressor ($R_{Agg2}$), and thus, the voltage transition on the column becomes 
\{GND$\rightarrow$VDD/2$\rightarrow$VDD$\rightarrow$VDD/2$\cdots$\}.

\begin{figure}[ht]
    \centering
    \includegraphics[width=\linewidth]{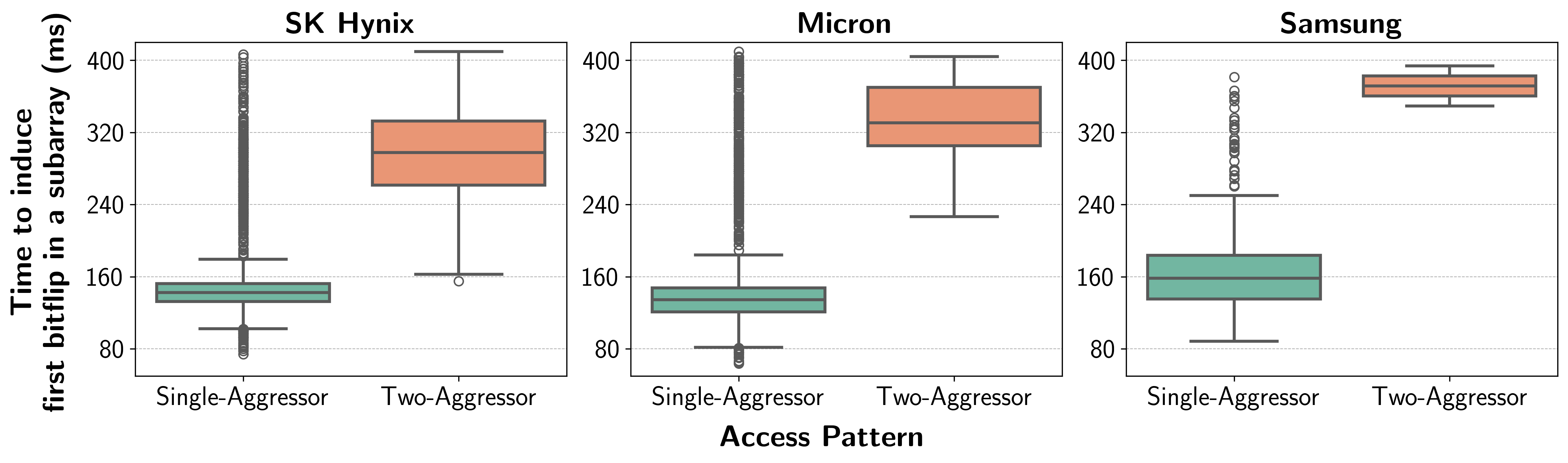}
    \caption{\ieycr{0}{Distribution of t}ime to induce \ieycr{0}{the} first \X{} bitflip \ieycr{0}{for two different access patterns.}}
    \label{fig:access_patt}
\end{figure}

\figref{fig:access_patt} shows the time to induce the first \X{} bitflip for \omcr{0}{the} two access patterns. In both experiments, victim rows are initialized with all-1
\ieycr{0}{as we observe only 1 to 0 bitflips for \X{} (\secref{sec:bitflip_direction})}.

\observation{Single-aggressor access pattern induces the first \X{} bitflip in a subarray significantly faster than the two-aggressor access pattern.}
For SK Hynix, Micron, and Samsung modules, the single-aggressor access pattern induces the first \X{} bitflip \param{1.83}×, \param{1.92}×, and \param{2.16}× faster than the two-aggressor access pattern, respectively.

\ieycr{0}{This observation shows that toggling a column through the sequence of GND$\rightarrow$VDD/2$\rightarrow$VDD (i.e., the two-aggressor access pattern) is less effective than toggling it only between VDD/2 and GND (i.e., the single-aggressor access pattern). }We hypothesize that this is because \ieycr{0}{the two-aggressor access pattern} results in a higher average column voltage than \ieycr{0}{the single-aggressor access pattern, resulting in lower \X{} vulnerability in DRAM chips, i.e., longer time to induce the first \X{} bitflip} (\secref{sec:avg_bl_voltage}).

\takeaway{\ieycr{1}{Access pattern greatly affects a DRAM chip’s vulnerability to \X{}.}}

\subsection{Data Pattern}
\label{sec:id_dp}
We analyze the effect of the aggressor \ieycr{1}{and victim} data patterns using five \ieycr{1}{aggressor and victim data pattern pairs. In this experiment, victim cells are initialized with the negated aggressor data pattern (e.g., if the aggressor data pattern is all-0, the victim data pattern is all-1)}. We omit the inverse of \ieycr{0}{the} tested patterns as either we 1) observe almost the same trend or 2) study it before (all-1 \ieycr{0}{in \secref{sec:dp_column}}).

\begin{figure}[ht]
    \centering
    \includegraphics[width=\linewidth]{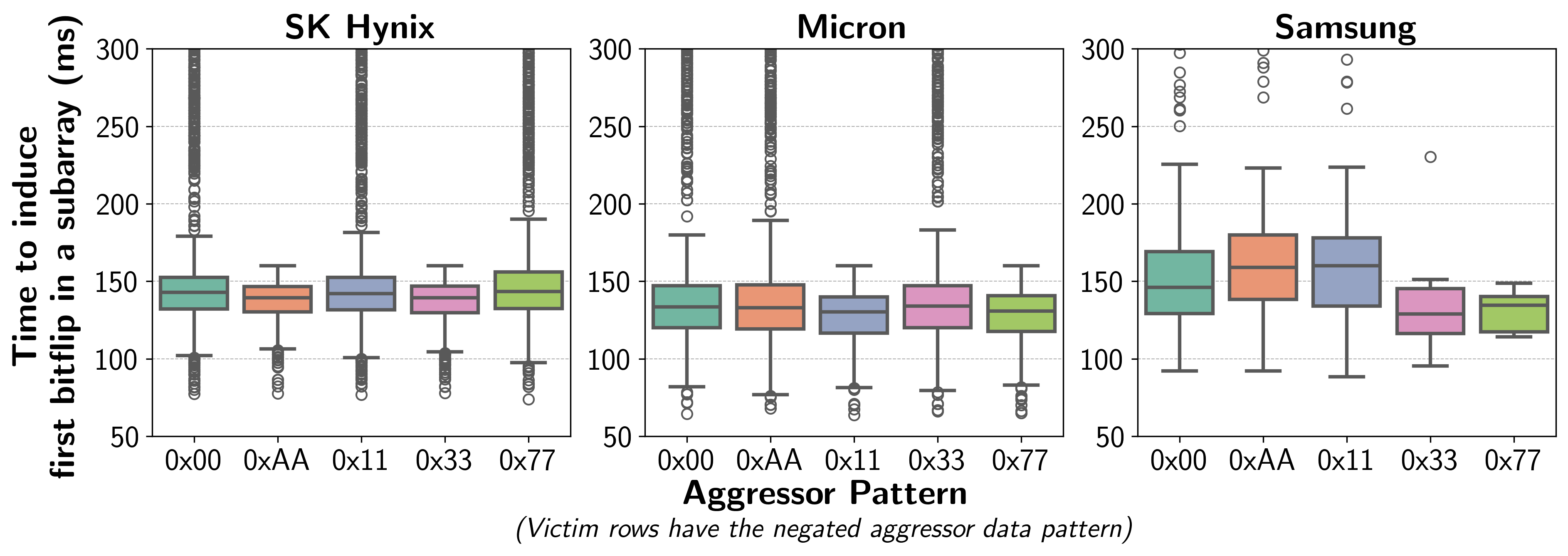}
    \caption{\ieycr{0}{Distribution of t}ime to induce the first \X{} bitflip \ieycr{0}{for five different} \ieycr{1}{aggressor and victim} data pattern \ieycr{1}{pairs}.}
    \label{fig:ch_dp}
\end{figure}

\noindent\textbf{Time to Induce the First \X{} Bitflip.}
\figref{fig:ch_dp} shows the distribution of time to induce the first \X{} bitflip in a subarray for \omcr{0}{five} data patterns. 

\observation{Data pattern has a small effect on the time to induce \omcr{0}{the} first \X{} bitflip \omcr{0}{in a subarray}.}
Across all data patterns, \ieycr{0}{the} average time to induce the first \X{} bitflip varies \ieycr{0}{by} at most \param{1.31}x. 
We hypothesize that \ieycr{0}{this small variation (1.31x) suggests that bitline-to-bitline interference does not strongly affect the time to induce the first \X{} bitflip. This is because the weakest cell determines the time to induce the first \X{} bitflip. If the weakest cell is connected to a column written with logic-0, it flips at nearly the same time regardless of the values in neighboring columns/bitlines. }

\begin{figure}[ht]
    \centering
    \includegraphics[width=\linewidth]{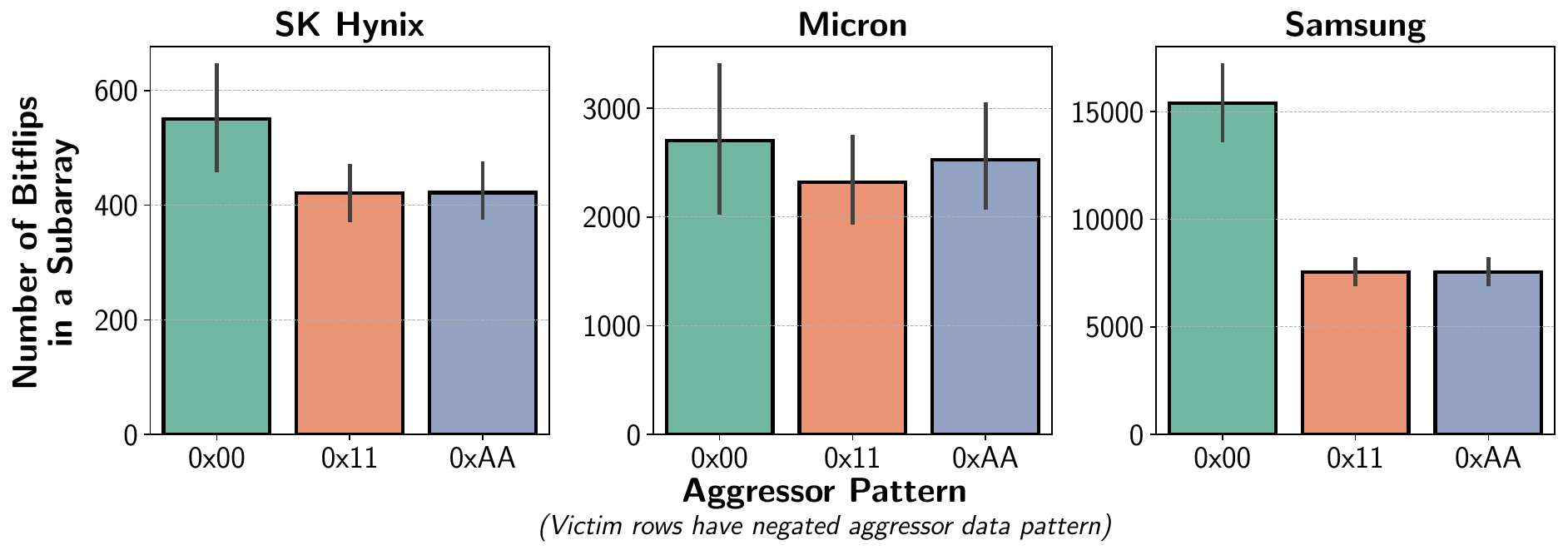}
    \caption{{Total \omcr{0}{number of} \X{} bitflip\omcr{0}{s} in a subarray \ieycr{0}{for three different} \ieycr{1}{aggressor and victim} data pattern \ieycr{1}{pairs}.}}
    \label{fig:ber_dp}
\end{figure}

\noindent\textbf{{Total \omcr{0}{Number of} Bitflips in a Subarray.}} {\figref{fig:ber_dp} shows the total \X{} bitflips in a subarray with a refresh interval of 512ms. \ieycr{0}{To maintain a reasonable experiment time, we test three different \ieycr{1}{aggressor and victim} data pattern \ieycr{1}{pairs}.}}

\observation{{Data pattern greatly affects total \X{} bitflips in a subarray: having more logic-0 \omcr{1}{values} in the \ieycr{0}{perturbed columns} generates more bitflips.}}

{For example, in Samsung chips, \omcr{1}{an aggressor data pattern of} 0x00 induces \param{2.04}x more \X{} bitflips than \omcr{1}{an aggressor data pattern of} 0xAA on average. We hypothesize that this is due to the bitflip direction characteristics of \X{}: only victim cells that are initialized with logic-1 experience bitflips \omcr{0}{(see \secref{sec:bitflip_direction})}. Thus, \ieycr{1}{since victim cells are initialized with the negated aggressor data pattern,} \omcr{0}{more} logic-0 \omcr{0}{values} in the aggressor pattern (which means more victim cells are initialized with logic-1) \omcr{0}{lead to} more \X{} bitflips.}

\subsection{{Location of the Aggressor Row in a Subarray}}
\label{sec:id_loc}
\begin{figure}[ht]
    \centering
    \includegraphics[width=\linewidth]{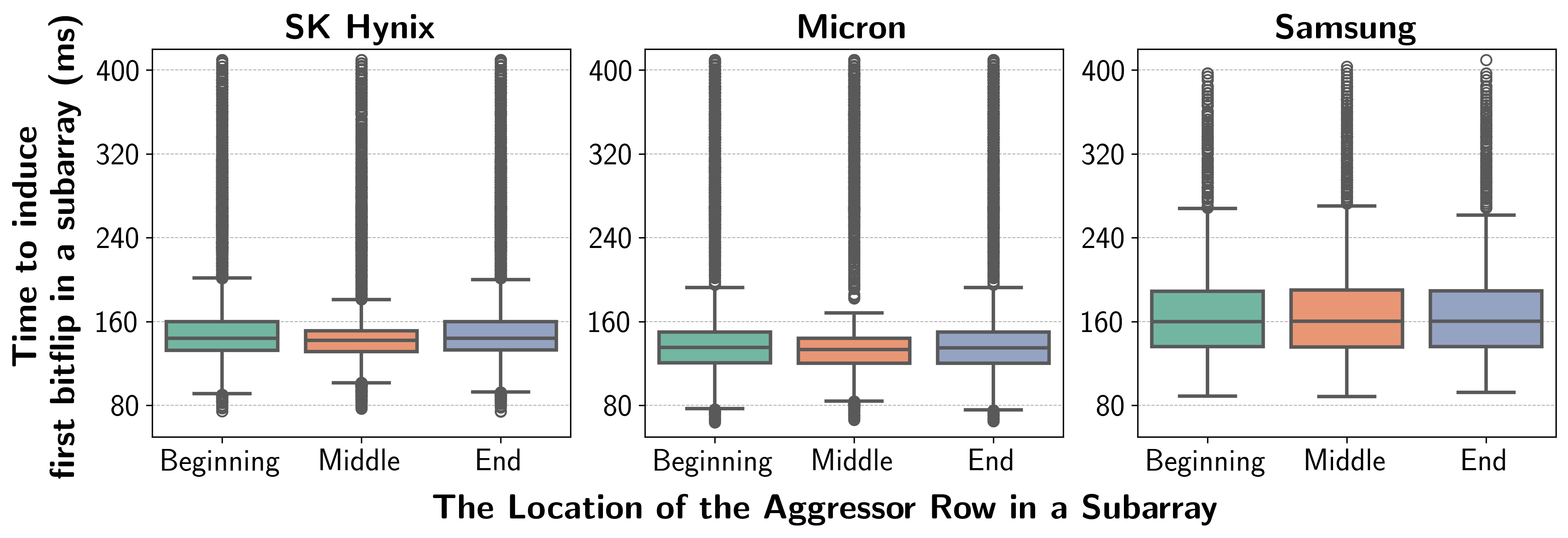}
    \caption{{Distribution of time to induce the first \X{} bitflip based on the aggressor row's location in a subarray.}}
    \label{fig:hcfirst_loc}
\end{figure}

{To understand the effects of spatial variation on \X{}, we analyze how the location of an aggressor row in a subarray affects the \X{} vulnerability. To do so, we test three aggressor row locations: \one{} "Beginning": the first row of the subarray, \two{} "Middle": the middle row in the subarray, and \three{} "End": the last row of the subarray. 
\figref{fig:hcfirst_loc} shows the distribution of time to induce the first bitflip in a subarray across DRAM subarrays (y-axis) based on \ieycr{0}{the} aggressor row’s location in a subarray (x-axis).} 

\observation{{The location of the aggressor row \omcr{0}{only} slightly affects the time to induce the first \X{} bitflip.}}
{We observe that across all tested manufacturers, there is at most \param{1.08}x variation on average when the location of the aggressor row in a subarray changes.}

\subsection{Effectiveness of System-Level ECC}
\label{sec:id_ecc}
\begin{figure}[ht]
    \centering
    \includegraphics[width=\linewidth]{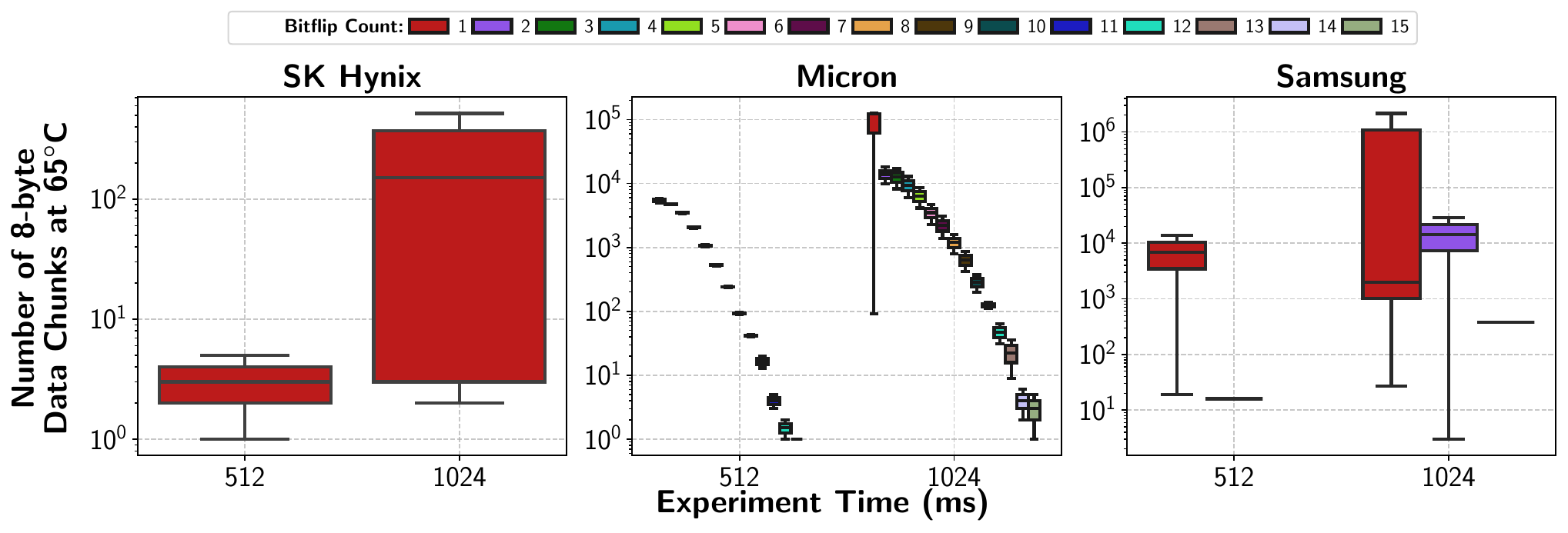}
    \caption{Distribution of 8-byte data chunks with different \X{} bitflip counts.}
    \label{fig:ecc}
\end{figure}

\ieycr{3}{A} system that uses Error-Correcting Codes (ECC)~\cite{dell1997white,gong2018memory,hamming1950error,kang2014co,meza2015revisiting,micron2017whitepaper,nair2016xed,patel2019understanding,bianca-sigmetrics09,dram-field-analysis3,patel2020beer,patel2021harp} can potentially protect against \X{} bitflips if those bitflips are distributed \atb{across the DRAM chip} such that no ECC codeword contains more bitflips than ECC can correct.
\figref{fig:ecc} shows the distribution of \X{} bitflips across 8-byte data chunks for all tested DRAM modules for 512ms and 1024ms \omcr{0}{refresh intervals}. We use 8-byte data chunks as DRAM ECC typically uses 8-byte or larger datawords~\cite{im2016im,kwak2017an,micron2017whitepaper,oh20153,son2015cidra,mineshphd,patel2019understanding,patel2020beer,patel2021harp,patel2022case}.

\observation{\X{} induces more bitflips than today's widely-used ECCs can correct or even detect.}
We observe that many 8-byte data chunks experience 3 (up to 15) \X{} bitflips, which the SECDED ECC~\cite{im2016im,kwak2017an,micron2017whitepaper,oh20153,son2015cidra} \emph{cannot} correct or detect, in Micron and Samsung \omcr{0}{chips}.

\observation{Relying \emph{only} on ECC to mitigate all \X{}-induced errors is likely \omcr{0}{very} costly.}
Since \X{} can induce 15 bitflips in multiple 8-byte data chunks\omcr{0}{, a} (7,4) Hamming code~\cite{hamming1950error} could
correct such bitflips with very large DRAM storage overheads (75\%, \omcr{0}{i.e.,} three parity bits for every four data bits). Thus, relying on ECC
alone to prevent \X{} bitflips is likely a very expensive solution.

\noindent
\textbf{Effectiveness of On-Die ECC.}
{We evaluate the effectiveness of a (136,128) single-error-correcting (SEC) Hamming code~\cite{hamming1950error} (possibly implemented in DDR5 chips today~\cite{mineshphd,alam2022comet,kim2023unity,patel2019understanding,patel2020beer} for area efficiency and low cost overhead).\footnote{{The exact design\omcr{0}{s} of on-die ECC are kept strictly confidential by DRAM manufacturers\omcr{0}{~\cite{patel2024rethinking, patel2021harp}} and may be different from what we evaluate.}} This code corrects any one bitflip in a 136-bit codeword. The observed bitflip distributions (\figref{fig:ecc}) for Micron and Samsung already exceed the error correction capabilities of such codes.
Moreover, the SEC code can miscorrect codewords that have more than one bitflip~\cite{alam2022comet,cha2017defect,kim2023unity}. A miscorrection manifests as additional bitflip\ieycr{0}{(s)} in the codeword (i.e., the SEC code induces another bitflip by attempting to correct the erroneous codeword). We randomly generate 10K erroneous codewords with two bitflips (bitflip indices in the codeword are determined using a uniform random number generator). The evaluated SEC code miscorrects 88.5\% of the codewords, transforming a codeword with two bitflips into another with three bitflips.}

\observation{{Low-cost on-DRAM-die single-error-correcting codes \emph{cannot} correct all ColumnDisturb bitflips and are likely to induce additional bitflips in an erroneous codeword that has as few as only two bitflips.}}

\takeaway{Conventional DRAM ECC cannot protect against \X{} bitflips\ieycr{0}{,} and an ECC scheme that can protect against \X{} requires large overheads.}

\section{Implications}
\label{sec:implications}
\atb{Our findings have strong implications for \atbcr{0}{both} 1) the robustness of \emph{future} 
systems and 2) the system speedup and DRAM
energy saving benefits of \emph{existing} retention-aware \atbcr{0}{heterogeneous} refresh mechanisms
(e.g.,\ieycr{1}{~\retentionMechanisms{}}). 
\X{} could jeopardize the safety, security, and reliability of future computing systems \atbcr{0}{as more \X{} bitflips may manifest in the standard refresh window due to continuously shrinking DRAM technology node size}.
\secref{sec:implications-for-future-systems} \atbcr{0}{describe\atbcr{1}{s} and evaluate\atbcr{1}{s}} \param{two}
techniques that could mitigate \X{} bitflips at varying performance, energy, and area overheads. 
In light of our new
findings, we evaluate a retention-aware heterogeneous refresh mechanism \atbcr{0}{(RAIDR~\cite{liu2012raidr})} and 
show that its benefits over the
baseline DRAM periodic refresh dramatically decrease (e.g., by \param{53}\%,
see~\secref{sec:retention-aware-refresh-evaluation}) because 
at long refresh windows a large majority of DRAM rows
(see \secref{sec:blast_radius} and \secref{sec:id_temp}) exhibit \X{} bitflips.}

\subsection{\atb{Implications for System Robustness}}
\label{sec:implications-for-future-systems}
\atb{Observations
{1, 2, 4, 13, 17} and \param{19} demonstrate that \X{} induces
bitflips in victim rows that are thousands of DRAM rows away from, and even in a
different subarray \omcr{1}{from} the aggressor row.} {For example, the first bitflip we observe in 63.6ms is 374 rows away from the aggressor row. \omcr{1}{Unfortunately}, current RowHammer mitigations only refresh up to 8 \omcr{1}{neighboring} rows~\ieycr{1}{\mitigatingRowHammerAllCitations{}} as RowHammer \omcr{1}{or} RowPress bitflips induce bitflips in immediate neighbors of the aggressor rows~\cite{orosa2021deeper,kim2020revisiting,kim2014flipping, yaglikci2022understanding, lim2017active, park2016statistical, park2016experiments, ryu2017overcoming, yun2018study, lim2018study, luo2023rowpress, lang2023blaster, yaglikci2024svard, nam2024dramscope, olgun2023hbm,nam2023xray,luo2025revisiting,tugrul2025understanding,yuksel2025pudhammer,he2023whistleblower,luo2024experimental,olgun2025variable}. Simply modifying commercial or many other proposed read disturbance mitigations to preventively refresh all potential victim \omcr{1}{rows} (\omcr{1}{e.g.,} 3072 such victims) of \X{} \atbcr{0}{would induce prohibitive system performance and energy overheads due to the latency and energy cost of performing thousands of} preventive refresh \atbcr{0}{operations}.
For example, according to the latest DDR5 JEDEC Standard~\cite{jedec2024jesd795c}, refreshing +/-4 physically adjacent neighboring rows incurs 2x more latency \atbcr{0}{(tDRFMab = 560ns)} than refreshing +/-2 \atbcr{0}{(tDRFMab = 280ns}). Refreshing all rows in three consecutive subarrays (\omcr{1}{e.g.,} 3072 rows) would take an estimated, prohibitive, $\sim$\atbcr{1}{215}$\mu$s latency.\footnote{\ieycr{2}{Assuming 1) refreshing eight rows takes 560 ns~\cite{jedec2024jesd795c}, which means 70 ns to refresh one row and 2) rows across three subarrays are refreshed sequentially. In this case, refreshing 3072 rows requires $70 \times 3072 = 215\mu$s.}}
\atbcr{0}{Moreover, as DRAM chip
density increases, so does the number of rows in a subarray.
Therefore, securely mitigating \X{} bitflips would likely require refreshing 
a few thousand rows before an aggressor row is hammered or pressed to \atbcr{1}{the} point of failure\ieycr{1}{. However,} \emph{reactively} refreshing that many rows 
\ieycr{1}{incurs prohibitive memory access latencies.}
We describe and evaluate two solutions for \X{} that  \emph{proactively} refresh \omcr{1}{potential victim rows of} \X{} without inducing prohibitive memory access latenc\omcr{1}{ies}.}

\noindent
\textbf{{Increasing DRAM Refresh Rate}.}
\atb{A straightforward (and high-overhead) solution for \X{} is to \omcr{1}{indiscriminately} {increase} the DRAM refresh {rate}.}
\atb{DRAM periodic refresh would mitigate \X{}-induced bitflips 
\omcr{1}{if} the refresh period is set adequately short. The benefit of this solution is 
that it can be applied immediately in today's systems.
However, as many prior works~\retentionMechanisms{} show, \omcr{1}{increasing the refresh rate degrades} system performance and energy \omcr{1}{efficiency} 
because more frequent refresh operations 1) \omcr{1}{increase memory access latency}, 
2) \omcr{1}{reduce row buffer hit rate}, 3) \omcr{1}{degrade bank-level parallelism, and 4) 
cause more activity on the memory bus and in DRAM banks}. For example, \ieycr{2}{for a 32~Gb DDR5 chip~\cite{micronddr5,jedec2020ddr5},} reducing the default 
\atbcr{1}{(all bank)} refresh period \atbcr{1}{(REFab)} from \SI{32}{\milli\second} to \SI{8}{\milli\second} {(i.e., 4x higher refresh rate)} increases
the DRAM throughput loss due to periodic refresh operations from 10.5\% to
42.1\%\footnote{\atbcr{1}{DDR5 all bank refresh command latency (tRFC) is \SI{410}{\nano\second} for 32~Gb chip density~\cite{jedec2020ddr5}. A DRAM chip \emph{cannot} serve any memory request (read or write) for tRFC after every refresh command, causing throughput loss. The memory controller issues refresh commands more frequently as refresh period reduces: once every \SI{3.9}{\micro\second} and \SI{975}{\nano\second} respectively for a refresh period of \SI{32}{\milli\second} and \SI{8}{\milli\second}.}} and relative energy consumed
by refresh operations for an otherwise idle DRAM chip increases from
25.1\% to 67.5\%.}\footnote{\atbcr{1}{We estimate refresh and idle energy from manufacturer provided DRAM IDD current values~\cite{micronddr5}.}}

\atb{We believe that \omcr{1}{indiscriminately} {increasing} the DRAM refresh
{rate} to mitigate \X{}-induced bitflips is 
\emph{not} a performance- and energy-efficient
solution. We describe a more intelligent mechanism that
selectively and timely refreshes \emph{only} the victim
rows that are disturbed by \X{}.}

\noindent
\textbf{Proactively Refreshing \X{} Victim Rows \omcr{1}{(PRVR)}.}
\atb{We propose proactively refreshing \X{} victim rows to mitigate \X{} bitflips.
The key idea of PRVR is to refresh \emph{only}
the victim DRAM rows in three subarrays: the aggressor DRAM row's subarray and 
its two neighboring subarrays. PRVR refreshes \atbcr{0}{\emph{each} of the} $N$ victim DRAM rows across three subarrays \atbcr{0}{(e.g., for subarray size = 1024, $N$ = 3072)}
\emph{once} before the aggressor row is hammered or pressed enough times to induce a bitflip.
PRVR distributes the refresh operations targeting these N rows over the time it takes to induce 
the first bitflip in a subarray using \X{}, similar to how a periodic refresh command
refreshes a subset of all DRAM rows in a DRAM chip.}

\atb{PRVR is a higher-performance and more energy-efficient solution
than using a shorter DRAM refresh period. Assuming a default refresh period of \SI{32}{\milli\second},
a time to induce the first \X{} bitflip of \SI{8}{\milli\second}, and $N=3072$,
we analytically estimate PRVR's DRAM throughput and energy benefits
over using a fixed, \SI{8}{\milli\second} refresh period. 
PRVR reduces \SI{8}{\milli\second}
refresh period's throughput loss by \atbcr{1}{70.5}\% and refresh energy consumption by 73.8\%
assuming one DRAM row from each DRAM bank
is \atbcr{1}{continuously} hammered \atbcr{1}{to induce ColumnDisturb bitflips} in a 32 Gb DDR5 chip.}\footnote{\atbcr{1}{System integration and rigorous evaluation of PRVR (e.g., using worst-case ColumnDisturb access patterns) is out of this paper's scope. We expect future work to build on the key idea of PRVR and enable new efficient and effective ColumnDisturb mitigation mechanisms.}}
\atb{We conclude that proactively refreshing \X{} victim rows could
prevent \X{} bitflips at much smaller performance and energy
overheads than simply reducing the DRAM refresh period.}

\subsection{\atb{\omcr{2}{Implications} on Retention-Aware Refresh}}
\label{sec:retention-aware-refresh-evaluation}
\atb{Retention-aware heterogeneous refresh mechanisms\ieycr{1}{~\retentionMechanisms{}} leverage the heterogeneity
in DRAM cell data retention time to \omcr{1}{reduce} the number of DRAM row refresh
operations while maintaining data integrity. These mechanisms typically
classify a DRAM row as \emph{weak} if any of its cells exhibit a retention
failure during a relatively
short refresh window (e.g., \SI{64}{\milli\second}).
In contrast, a DRAM row that reliably retains data 
over a much longer time window (e.g., \SI{1024}{\milli\second})
is considered \emph{strong}. Consequently, the greater the proportion of strong rows 
in a DRAM chip, \omcr{1}{the higher the performance and energy 
reduction provided by a retention-aware heterogeneous refresh mechanism}. \X{} greatly reduces the benefits of
retention-aware heterogeneous refresh mechanisms because \X{} renders many more rows weak 
than retention failures alone at
a given refresh window, as observations \param{6, 18,} and~\param{19} demonstrate.}

\atb{\figref{fig:num-refreshes-vs-weak-rows} shows the number of 
DRAM row refresh operations (normalized to the number of refresh
operations performed by DDR4 \SI{64}{\milli\second} periodic refresh) 
\omcr{1}{as the} proportion
of weak rows in a chip \omcr{1}{varies}. We plot four lines, one each for 
four strong row retention time values:
\SI{128}{\milli\second}, \SI{256}{\milli\second},
\SI{512}{\milli\second}, and \SI{1024}{\milli\second}.}
\atb{A circle, diamond, and a square on a line show the 
number of refresh operations needed for empirically observed
average proportion of retention-weak rows across all tested modules,
average proportion of ColumnDisturb-weak rows across all tested modules,
and the maximum proportion of ColumnDisturb-weak rows across all tested modules, respectively.
A smaller y-axis value indicates smaller expected retention-aware heterogeneous refresh mechanism
speedup and energy benefits.}

\begin{figure}[!ht]
    \centering
    \includegraphics[width=1.0\linewidth]{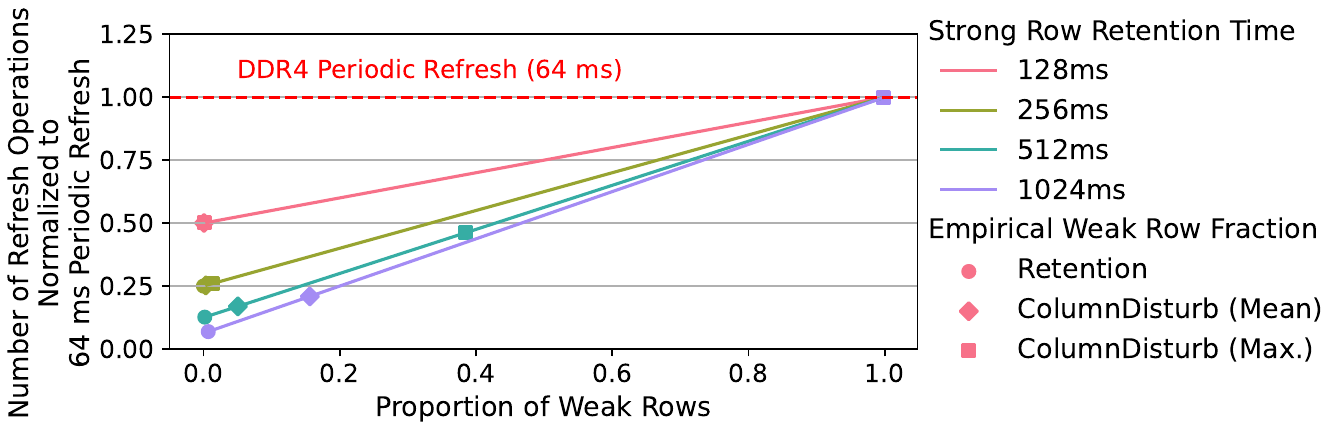}
    \caption{Number of DRAM row refresh operations needed. \atbcr{1}{We plot one line each for \atbcr{1}{four} strong row retention times \atbcr{1}{of 128, 256, 512, and 1024 milliseconds}.}}
    \label{fig:num-refreshes-vs-weak-rows}
\end{figure}

\atb{We make two key observations.
First, a relatively large strong row retention time can significantly
increase the expected speedup and energy benefits of retention-aware
refresh mechanisms. For example, for the empirically observed average 
retention-weak row proportion (circles), number of refresh operations
reduce\ieycr{1}{s} by 43.1\% when a strong row retention time of
\SI{1024}{\milli\second} is used instead of a strong row
retention time of \SI{128}{\milli\second}.
Second, ColumnDisturb significantly reduces the
expected speedup and energy benefits of retention-aware heterogeneous refresh mechanisms,
especially at high strong row retention times where
the empirical proportion of ColumnDisturb-weak rows greatly
exceed\ieycr{1}{s} that of retention-weak rows. For example, \ieycr{1}{in the presence of \X{}\omcr{2}{,}}
for a strong row retention time of \SI{1024}{\milli\second}, \omcr{2}{the} number of refresh operations increases by \param{3.02}$\times{}$ 
(purple diamond in~\figref{fig:num-refreshes-vs-weak-rows}) on average 
and by up to \param{14.43}$\times{}$ 
(purple square in~\figref{fig:num-refreshes-vs-weak-rows})\ieycr{1}{, compared to a baseline case without \X{} (i.e., only retention failures are present)}  across all tested modules.}

\takeaway{ColumnDisturb significantly degrades the expected speedup and energy
saving benefits of retention-aware heterogeneous refresh mechanisms.}
\begin{figure*}[tb]
    \centering
    \includegraphics[width=0.95\linewidth]{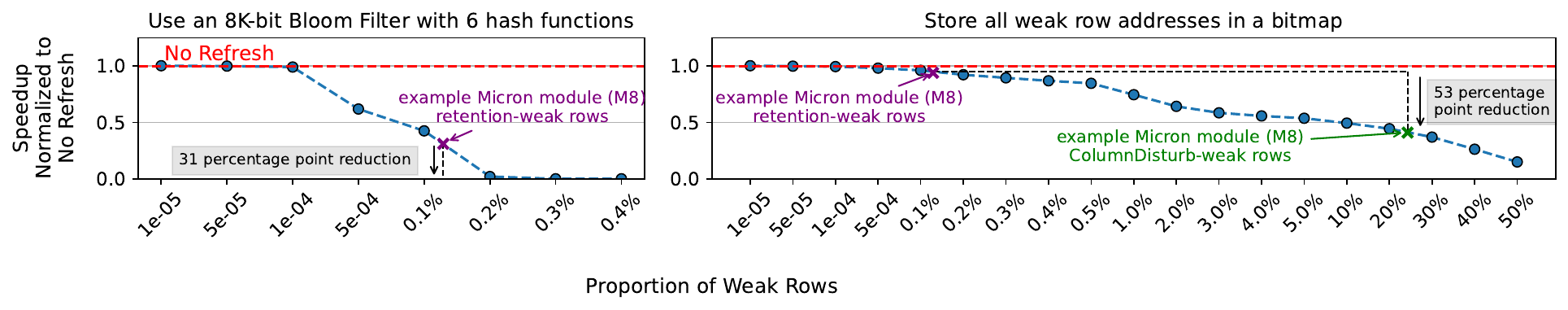}
    \caption{Speedup provided by RAIDR~\cite{liu2012raidr} over
    the proportion of weak rows in a DRAM module. Left and right subplots shows speedup for
    a low-area-overhead and high-area-overhead implementation, respectively. }
    \label{fig:performance}
\end{figure*}
\atb{To quantitatively demonstrate the speedup and energy benefits of 
retention-aware heterogeneous refresh mechanisms, we evaluate a state-of-the-art retention-aware 
refresh mechanism, RAIDR~\cite{liu2012raidr}, building on
the Self-Managing DRAM (SMD) framework~\cite{hassan2024self,hassan2024self.github}.\footnote{\atb{We 
use Ramulator~\cite{kim2016ramulator,ramulatorgithub,luo2023ramulator2,ramulator2github} and simulate
a modern computing system (configured
as the baseline system described \omcr{1}{in} Table 1 \omcr{1}{of}~\cite{hassan2024self}).
We use 20 highly-memory-intensive multi-programmed workload mixes,
each comprised of four single-core workloads each with last-level cache 
misses-per-kilo-instructions $\geq{}\!$ 10. 
We warm-up the caches by fast-forwarding 100 million (M) instructions. 
We run each multi-core simulation 
until each core executes at least 500M instructions.}}}
\atb{We evaluate two variants of RAIDR where 1) we store weak row addresses
in a space-efficient (low-area-overhead) Bloom Filter~\cite{bloom1970space,cohen03,bonomi2006improved} sized 8Kb with 6 hash functions and
2) we store weak row addresses in a space-inefficient 
(high-area-overhead) bitmap\footnote{\atb{Storing weak row 
addresses in a bitmap
allows the retention-aware heterogeneous refresh mechanism
to identify all weak rows accurately~\cite{qureshi2015avatar}. Compared to the Bloom Filter,
using a bitmap has higher weak row identification accuracy but incurs higher storage overheads.}} indexed using
the row address (1 bit per DRAM row, 2 Mb for a 16 GiB DDR4 module).
We refresh
a weak row every \SI{64}{\milli\second} and a strong row every \SI{1024}{\milli\second}.
We evaluate a hypothetical DRAM
system configuration called \omcr{1}{\emph{No Refresh}} where we do \emph{not} issue
any refresh operations to show the headroom for system speedup
and energy consumption improvements of retention-aware heterogeneous refresh mechanisms.}

\atb{\figref{fig:performance} shows the system speedup provided by
retention-aware refresh (in weighted speedup) on average across all 20 four-core workloads
normalized to the speedup of No Refresh. The left and right subplots,
respectively, show the performance \omcr{1}{of} the RAIDR variant that uses a Bloom Filter
and the RAIDR variant that stores each weak row's address in a bitmap. The x-axis shows
the evaluated proportion of weak row values in the simulated DRAM system.
We annotate the speedup of retention-aware refresh for empirically observed proportion of
1) retention-weak rows (purple $\times{}$) and 2) ColumnDisturb-weak rows (green $\times{}$)
in one Micron DDR4 module.}

\atb{We make two key observations. First, \X{} can completely negate the 
performance and energy \omcr{1}{(not shown in the figure)} benefits of space-efficient (using a Bloom Filter)
implementations of retention-aware heterogeneous refresh mechanisms. 
From \figref{fig:performance} (left)
we observe that as the proportion of weak rows increases from 1.00e-04 to 0.20\% (by 20$\times{}$),
the speedup and energy benefits reduce to a point where they are
almost completely \omcr{1}{eliminated} ($\approx{}\!$99 percentage point speedup and energy benefit reduction).
The 20$\times{}\!$ increase in the proportion of weak rows is well within
\ieycr{1}{our empirical observations: across all tested modules, the proportion of weak rows increased by up to \param{198}$\times{}$ (\figref{fig:blast_temp}).}

As a concrete example, for the annotated Micron module (purple $\times{}$), the speedup
and energy consumption benefits reduce by 31 percentage points and 32 percentage points (not shown), respectively,
as the proportion of weak rows increase\omcr{1}{s} to accommodate \X{}-weak rows.  
Second, \X{} can greatly reduce the speedup and energy benefits even of
space-inefficient (using a bitmap) implementations of retention-aware heterogeneous refresh mechanisms.
From \figref{fig:performance} (right), we observe that for the
annotated Micron DDR4 module, the speedup and energy consumption benefits reduce by 53 and 50 percentage
points (not shown), respectively.}

\takeaway{\X{} can completely negate the performance and energy benefits
of low-area-overhead retention-aware heterogeneous refresh mechanisms. \X{} greatly
reduces the benefits of high-area-overhead retention-aware heterogeneous refresh mechanisms.}

\atb{We conclude that the proportion of strong DRAM rows (that can endure
high retention times without a bitflip) for a given refresh window significantly reduces with \X{}. 
This reduction greatly hinders (and can completely \omcr{1}{eliminate}) the performance and energy benefits of retention-aware heterogeneous
refresh mechanisms.}

\section{Related Work}
To our knowledge, this is the first work to experimentally demonstrate and characterize \X{}, a column-based read-disturb phenomenon in real DRAM chips.

\noindent\textbf{DRAM Read Disturbance Characterization.} Many works~\cite{orosa2021deeper,kim2020revisiting,kim2014flipping, yaglikci2022understanding, lim2017active, park2016statistical, park2016experiments, ryu2017overcoming, yun2018study, lim2018study, luo2023rowpress, lang2023blaster, yaglikci2024svard, nam2024dramscope, olgun2023hbm,nam2023xray,luo2025revisiting,tugrul2025understanding,yuksel2025pudhammer,he2023whistleblower,luo2024experimental,olgun2025variable} experimentally demonstrate how a real DRAM chip's read disturbance \omcr{0}{vulnerability} varies with 1)~DRAM refresh rate~\cite{hassan2021utrr,frigo2020trrespass,kim2014flipping}, 2)~the distance between aggressor and victim rows~\cite{kim2014flipping,kim2020revisiting,lang2023blaster}, 3)~DRAM generation and technology node~\cite{orosa2021deeper,kim2014flipping,kim2020revisiting,hassan2021utrr}, 4)~temperature~\cite{orosa2021deeper,park2016experiments}, 5)~time the aggressor row stays active~\cite{orosa2021deeper,park2016experiments,olgun2023hbm,olgun2024read,luo2023rowpress,nam2024dramscope,nam2023xray,luo2025revisiting,luo2024rowpress,luo2024experimental}, ~6)~location of the victim cell~\cite{orosa2021deeper,olgun2023hbm,olgun2024read,yaglikci2024svard}, 7)~wordline voltage~\cite{yaglikci2022understanding}, 8)~supply voltage~\cite{he2023whistleblower}, \ieycr{1}{9) reduced DRAM timing
parameters~\cite{tugrul2025understanding, yuksel2025pudhammer}, and 10)
time~\cite{olgun2025variable}.} None of these works analyze or demonstrate \X{}, a column-based read disturbance phenomenon.

\noindent\textbf{DRAM Retention.} Many works experimentally study DRAM data retention~\cite{jung2015omitting,khan2014efficacy,khan2016parbor,khan2016case,kim2015architectural,liu2013experimental,meza2015revisiting,bianca-sigmetrics09,weis2015retention,patel2017reaper,gong2018memory}. 
Several works~\retentionMechanisms{} present various error profiling algorithms to identify data retention failures when reducing the refresh rate and propose retention-aware heterogeneous refresh mechanisms to alleviate refresh overheads. However, we show that \X{} can significantly amplify the bitflips across subarrays and can cause significant performance and area overheads to retention-aware heterogeneous refresh mechanisms.

\noindent\textbf{Bitline Disturbance in 4F$^2$ VCT DRAM.} \hluo{Emerging high-density 4F$^2$ DRAM architectures with Vertical Channel Transistors (VCT) are known to be vulnerable to disturbances from the bitline with a hammering-like access pattern~\cite{Chung2011Novel4F2, Min2020Vertical, Cho2018NovelBand, Cho2018Suppression, Liu2025BitLineHammering}. However, since the device architecture of 4F$^2$ VCT DRAM is completely different from the contemporary COTS 6F$^2$ DRAM that we characterize in this work, we believe the error mechanism they study (i.e., floating body effect~\cite{Cho2018NovelBand,Cho2018Suppression}) is different from the one(s) that causes \X{} (see \secref{sec:foundational} for our analyses and hypotheses).}

\section{Conclusion}
\ieycr{2}{We present the first experimental demonstration and analysis of a new read disturbance phenomenon, \X{}, in modern DRAM chips: hammering or pressing an aggressor row disturbs DRAM cells through a DRAM column and induces bitflips in DRAM cells sharing the same columns as the aggressor row (across as many as three DRAM subarrays).
Our experimental characterization of \nCHIPS{} real DDR4 and 4 HBM2 DRAM chips reveals that 
1) \X{} is fundamentally different from RowHammer \& RowPress,
2) \X{} worsens with DRAM technology scaling, 
3) introduces bitflips within the nominal refresh window under nominal operating conditions in some real DRAM chips, and
4) DRAM is significantly more vulnerable to \X{} than to retention failures. 
We describe and evaluate two \X{} mitigation techniques. We hope and believe that our detailed demonstration and analyses will inspire future works on better understanding \X{} at the device-level and mitigating it before it becomes a bigger vulnerability in future DRAM chips.}

\section*{Acknowledgments} 
\ieycr{0}{We thank the anonymous reviewers of {MICRO 2025} for feedback. {We thank the} SAFARI Research Group members (especially Konstantinos Sgouras and Harsh Songara) for
{constructive} feedback and the stimulating intellectual {environment.} We
acknowledge the generous gift funding provided by our industrial partners
({especially} Google, Huawei, Intel, Microsoft), which has been instrumental in
enabling the research we have been conducting on read disturbance in DRAM {in
particular and memory systems in
general~\cite{mutlu2023retrospective}.} This work was in part
supported by \omcr{1}{a} Google Security and Privacy Research Award and the Microsoft
Swiss Joint Research Center.}

\balance
\bibliographystyle{unsrt}
\bibliography{refs}

\end{document}